\documentclass[aps,prb,twocolumn,superscriptaddress,showpacs]{revtex4}
\usepackage{graphicx}
\usepackage{longtable}
\usepackage{array}
\usepackage{braket}
\newcolumntype{L}{>{\centering\arraybackslash}m{2cm}}
\newcolumntype{R}{>{\centering\arraybackslash}m{1.5cm}}
\newcolumntype{K}{>{\centering\arraybackslash}m{1.3cm}}

\def \DTO{Dy$_2$Ti$_2$O$_7$}
\def \HTO{Ho$_2$Ti$_2$O$_7$}

\begin{document}


\title{Neutron Spectroscopic Study of Crystalline Electric Field Excitations in Stoichiometric and Lightly Stuffed $\bf{Yb_2Ti_2O_7}$}

\author{J. Gaudet}
\email{gaudej@mcmaster.ca}
\affiliation{Department of Physics and Astronomy, McMaster University, Hamilton, ON L8S 4M1 Canada}

\author{D.D. Maharaj}
\affiliation{Department of Physics and Astronomy, McMaster University, Hamilton, ON L8S 4M1 Canada}

\author{G. Sala}
\affiliation{Department of Physics and Astronomy, McMaster University, Hamilton, ON L8S 4M1 Canada}

\author{E. Kermarrec}
\affiliation{Department of Physics and Astronomy, McMaster University, Hamilton, ON L8S 4M1 Canada}

\author{K. A. Ross}
\affiliation{Department of Physics, Colorado State University, Fort Collins, Colorado 80523-1875, USA}

\author{H. A. Dabkowska}

\affiliation{Brockhouse Institute for Materials Research, Hamilton, ON L8S 4M1 Canada}

\author{A.I. Kolesnikov}
\affiliation{Oak Ridge National Laboratory, Oak Ridge, Tennessee 37831-6473, USA}

\author{G. E. Granroth}
\affiliation{Oak Ridge National Laboratory, Oak Ridge, Tennessee 37831-6473, USA}

\author{B. D. Gaulin}
\affiliation{Department of Physics and Astronomy, McMaster University, Hamilton, ON L8S 4M1 Canada}
\affiliation{Brockhouse Institute for Materials Research, Hamilton, ON L8S 4M1 Canada}
\affiliation{Canadian Institute for Materials Research, 180 Dundas Street West, Toronto, Ontario M5G 1Z8, Canada}





\date{\today}

\begin{abstract}
Time-of-flight neutron spectroscopy has been used to determine the crystalline electric field (CEF) Hamiltonian, eigenvalues and eigenvectors appropriate to the $J$ = 7/2 Yb$^{3+}$ ion in the candidate quantum spin ice 
pyrochlore magnet $\rm Yb_2Ti_2O_7$.  The precise ground state (GS) of this exotic, geometrically-frustrated magnet is known to be sensitive to weak disorder associated with the growth of single crystals from the melt.  
Such materials display weak ``stuffing" wherein a small proportion, $\approx$ 2\%, of the non-magnetic Ti$^{4+}$ sites are occupied by excess Yb$^{3+}$.  We have carried out neutron spectroscopic 
measurements on a stoichiometric powder sample of Yb$_2$Ti$_2$O$_7$, as well as a crushed single crystal with weak stuffing and an approximate composition of Yb$_{2+x}$Ti$_{2-x}$O$_{7+y}$ with $x$ = 0.046.  
All samples display three CEF transitions out of the GS, and the GS doublet itself is identified as primarily composed of m$_J$ = $\pm$1/2, as expected.  
However,``stuffing" at low temperatures in Yb$_{2+x}$Ti$_{2-x}$O$_{7+y}$ induces a similar finite CEF lifetime as is induced in stoichiometric Yb$_2$Ti$_2$O$_7$ by elevated temperature.  
We conclude that an extended strain field exists about each local ``stuffed" site, which produces a distribution of random CEF environments in the lightly stuffed Yb$_{2+x}$Ti$_{2-x}$O$_{7+y}$, in addition 
to producing a small fraction of Yb-ions in defective environments with grossly different CEF eigenvalues and eigenvectors.
\end{abstract}



\maketitle



\section{INTRODUCTION}

Geometrically-frustrated magnetic materials are of great current interest due to the diversity of exotic ordered and disordered ground states (GSs) that they display~\cite{Lacroix}.  In particular, cubic pyrochlore magnets with chemical 
composition A$_2$B$_2$O$_7$ have been a playground for geometric frustration as both the A$^{3+}$ and B$^{4+}$ sites, independently, reside on interpenetrating networks of corner-sharing tetrahedra (see Fig.~1), 
one of the canonical architectures for frustrated ground states in three dimensions~\cite{Subramanian}.  The rare earth titanate pyrochlores have played a pivotal role in the development of the field, as Ti$^{4+}$ at the B site is 
non-magnetic, and the A site can be occupied by all trivalent rare earth ions from Sm$^{3+}$ to Lu$^{3+}$.  Many of these rare earth titanates therefore have a magnetic A sub-lattice, and the family as 
a whole gives rise to different combinations of magnetic anisotropies and interactions, which in turn are responsible for the diversity of exotic ground states~\cite{Greedan}.

\begin{figure}[h]
\includegraphics[width=9cm]{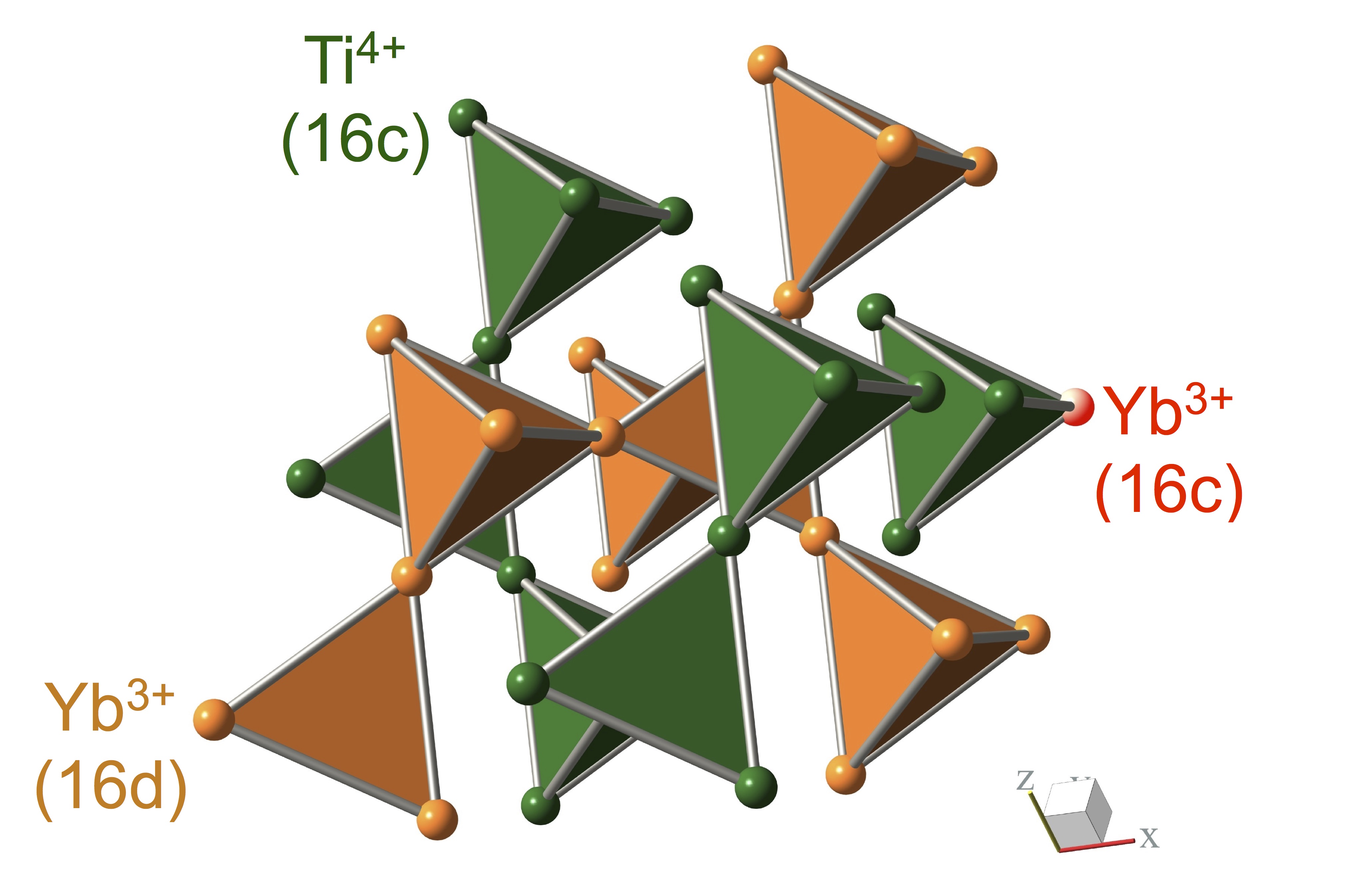}
\label{fig: Stuffing}
\caption{\emph{The stuffed pyrochlore structure of Yb$_{2}$Ti$_{2}$O$_{7}$. --} The pyrochlore lattice consists of interpenetrating networks of corner-sharing tetrahedra which are generated independently by the Yb$^{3+}$ (magnetic) and Ti$^{4+}$ (non-magnetic) cations. The pyrochlore lattice is said to be stuffed wherein a Yb$^{3+}$ ions, which are normally found on the 16d site (also called A-site), also occupy the Ti$^{4+}$ 16c site (B-site).}
\end{figure}

As an example, the classical spin ice state has been the focus of much attention~\cite{Bramwell2001,Castelnovo2008,DenHertog2000}, and it results from a combination of local Ising anisotropy~\cite{Rosenkranz1,Harris1997} and net ferromagnetic interactions~\cite{Harris1997} on the pyrochlore lattice, such that the Ising magnetic 
moments on each tetrahedron obey ``ice rules" with  two spins pointing into each tetrahedron and two spins pointing out, analogous to the water ice model proposed by Pauling~\cite{Pauling}.  This results in a sixfold degeneracy for a given tetrahedron, and a macroscopic degeneracy 
for the three dimensional network as a whole.  Yb$_2$Ti$_2$O$_7$ has attracted much recent attention ~\cite{Yasui2003,Chang2012,Hodges2002,Ross2009,Ross2011} as a candidate for a quantum spin ice ground state ~\cite{Hermele2004,GingrasRev}, wherein effective S = 1/2 degrees of freedom decorate a 
pyrochlore lattice and interact via net ferromagnetic interactions~\cite{Bramwell2000}.  The GS phase diagram and microscopic Hamiltonian appropriate to Yb$_2$Ti$_2$O$_7$ have been extensively studied~\cite{SavaryPRL,SavaryPRB}.  
The microscopic Hamiltonian itself has been determined by modelling spin wave dispersion and neutron intensities in the high magnetic field, low temperature state of Yb$_2$Ti$_2$O$_7$~\cite{RossPRX}. This work convincingly showed anisotropic exchange to be the relevant form of the interactions at low temperatures.\\

One of the most interesting features of the GS properties of $\rm Yb_2Ti_2O_7$ is its apparent sensitivity to small levels of defects that are present in real materials~\cite{RossStuffing,Lau2006,Lau2008}.  Stoichiometric $\rm Yb_2Ti_2O_7$ 
is known to display a large and sharp anomaly in its heat capacity, C$_p$, near $\sim$ 265 mK. However, this anomaly has been observed to be sample dependent with samples displaying broader anomalies at lower temperatures depending on the exact stoichiometry of the material studied ~\cite{Takatsu2012,Yaouanc2011,PalmasPhysicaB,Blote,Yaouanc2011B}.  This phenomena is very unusual for a three dimensional magnet, as the defect levels involved are at the limits of detectability 
by conventional techniques.  The variation in stoichiometry, where characterized, is on the order of 1$\%$ level, far removed from percolation thresholds in three dimensions.
 
Polycrystalline samples of $\rm Yb_2Ti_2O_7$ tend to display the sharpest and highest temperature C$_p$ anomalies, likely due to the lower temperatures required for their synthesis, leading to less TiO$_2$ volatization.  Single crystals grown by floating zone image furnace techniques typically display broad low temperature C$_P$ anomalies, or no anomalies, and sometimes show multiple peaks in C$_P$ at low temperature.  Ross \textit{et al.}~\cite{RossStuffing} undertook a detailed neutron crystallographic study of both powder and crushed single crystal $\rm Yb_2Ti_2O_7$ samples, representative of those that displayed sharp and broad low temperature C$_P$  signatures, respectively, and showed that the crushed single crystal sample displayed weak ``stuffing": a structural defect in which a slight excess of 
Yb occupies the Ti sub-lattice.  Stuffing is illustrated schematically in Fig.~1, wherein Yb$^{3+}$ ions occupy both the 16d site of the $Fd\bar{3}m$ cubic space group, as well as act as impurities on the 16c site 
normally occupied by Ti$^{4+}$.  Weak stuffing was shown to occur at the 2.3$\%$ level in the crushed single crystals with composition Yb$_{2+x}$Ti$_{2-x}$O$_{7+y}$ grown by floating zone image techniques.   In contrast the powder sample grown by solid state synthesis was shown to be stoichiometric.
 
In this paper we report neutron spectroscopic measurements of the crystalline electric field (CEF) excitations in the two powder samples, the stoichiometric powder and the crushed single crystal with 2.3 $\%$ ``stuffing" 
previously studied by Ross et al.  These measurements allow us to accurately determine the eigenvalues and eigenvectors appropriate to the 4 doublets which make up the $J$ = 7/2 CEF manifold for Yb$^{3+}$ 
in Yb$_2$Ti$_2$O$_7$.  Measurements on both stoichiometric and lightly ``stuffed" samples allow us to investigate the role of ``stuffing" on the CEF levels associated with Yb$^{3+}$ ions properly residing 
on the A-site of the pyrochlore structure.  With these benchmark measurements in hand, we calculated CEF eigenvalues and eigenvectors for the ``stuffed" Yb$^{3+}$ ions residing on the B site, as well as for the 
A site Yb$^{3+}$ ions in the presence of oxygen vacancies ~\cite{Sala2014}.


\section{\label{sec: CEF1} Calculated Crystal Field Levels for $\bf{Yb^{3+}}$ at the A-site}

Hund's rules enable the determination of the total angular momentum $J$ of the $\rm Yb^{3+}$ ion. The electronic configuration of $\rm Yb^{3+}$ is $4f^{13}$, resulting in $J = \frac{7}{2}$ which is $2J+1 = 8$-fold degenerate. 
Within the pyrochlore structure this degeneracy is lifted by the CEFs at the Yb$^{3+}$ site due largely to the presence of the eight neighbouring O$^{2-}$ ions. As illustrated in Fig.~2 (left panel), the oxygen environment at
the A-site consists of a scalenohedron, which is a cube distorted along one diagonal that forms the local [111] axis. Six oxygen ions, commonly referred to as O(2), are located on a plane perpendicular to this direction, which is a three-fold rotation axis. The other two oxygen ions, referred to as O(1), are located along the local [111] axis in the geometric centre of the tetrahedra defined by the A-site $\rm Yb^{3+}$ ions. By contrast, the environment at the B-site is a trigonal anti prism made of six O(2) oxygen ions surrounding the transition metal, as shown in the right panel of Fig.~2.

\begin{figure}[h]
\includegraphics[width=8.8cm]{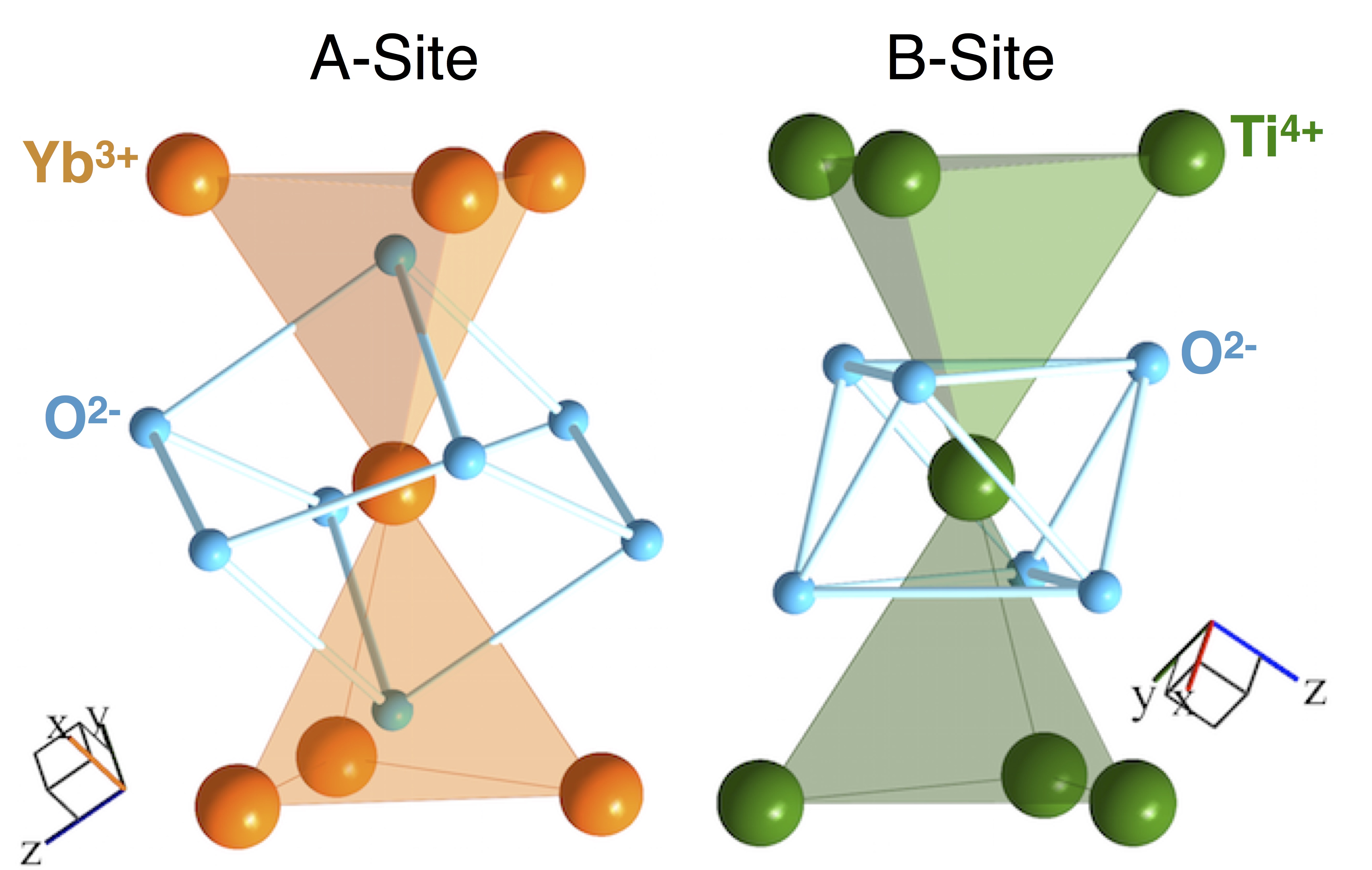}
\label{fig: YbEnv}
\caption{\emph{A comparison between the A-site and B-site oxygen environment in the pyrochlore structure of Yb$_{2}$Ti$_{2}$O$_{7}$. --} The left panel of the figure shows the scalenohedron environment generated by the oxygen ions at the A-site where the $\rm Yb^{3+}$ resides. The symmetry of this structure is similar to that at the B-site (right panel) where the $\rm Ti^{4+}$ ions are located.}
\end{figure}

Following Prather's convention~\cite{Prather}, the 3-fold axis should be placed along $\hat{z}$ of the reference system in order to minimize the number of CEF parameters in the Hamiltonian. Therefore the resulting CEF Hamiltonian for $\rm Yb^{3+}$ on the A-site can be written as:

\begin{eqnarray}
\mathcal{H}_{CEF}= B^0_2\hat{O}^0_2 + B^0_4\hat{O}^0_4 +  B^3_4\hat{O}^3_4 +
\nonumber
\\
B^0_6\hat{O}^0_6 + B^3_6\hat{O}^3_6 + B^6_6\hat{O}^6_6.
\label{eq: HCEF}
\end{eqnarray}

Here we employ the Steven's operators $\hat{O}^m_n$~~\cite{Stevens} and CEF parameters $B^m_n$ to approximate the Coulomb potential generated by the crystalline electric field due to the neighbouring oxygen atoms.

The determination of the crystal field parameters $B^m_n$ in Eq.~\ref{eq: HCEF} is well suited to inelastic neutron spectroscopy. The unpolarized neutron partial differential magnetic cross-section can be written within the dipole approximation as~\cite{Squires}:
\begin{eqnarray}
\frac{d^2\sigma}{d\Omega dE'} = C\frac{k_f}{k_i}F(|Q|)S(|Q|,\hbar \omega),
\end{eqnarray} 
where $\Omega$ is the scattered solid angle, $E'$ the final neutron energy, $\frac{k_f}{k_i}$ the ratio of the scattered and incident momentum of the neutron, C a constant and $F(|Q|)$ is the magnetic form factor. The scattering function $S(|Q|,\hbar\omega)$ gives the relative scattered intensity due to transitions between different CEF levels. At constant temperature and wave vector $|Q|$, we have:

\begin {eqnarray}
S(|Q|,\hbar\omega) = \sum_{i,i'}\frac{(\sum_{\alpha}  |\langle i {| J_{\alpha} | i'\rangle |}^2) \mathrm{e}^{-\beta E_{i}} }{\sum_j \mathrm{e}^{-\beta E_{j}}} L(\Delta E + \hbar \omega),
\end{eqnarray} 

where $\alpha = x,y,z$ and $L(\Delta E + \hbar \omega) = L(E_{i} - E_{i'} + \hbar \omega)$ is a Lorentzian function which ensures energy conservation as the neutron induces transitions between the CEF levels $i \rightarrow i'$, which possess a finite energy width or lifetime.

The procedure for fitting our inelastic neutron scattering (INS) data assumes an initial set of CEF Hamiltonian parameters. The Hamiltonian is then diagonalized to find the corresponding CEF eigenfunctions and eigenvalues. $S(|Q|,\hbar\omega$) is then computed and directly compared with the experimental results and, finally, the CEF parameters are tuned such that the $\chi^2$ between the calculated and measured $S(|Q|,\hbar\omega$) is minimized.

\section{EXPERIMENTAL DETAILS}

Now we turn our attention to the details of the INS experiment which was conducted on two samples of Yb$_2$Ti$_2$O$_7$ of different stoichiometry. We will refer to these two samples as the stoichiometric (x = 0) and stuffed powder (x = 0.046) samples hereafter. Specific details regarding their synthesis and characterization can be found in previous work by Ross \textit{et al.}~\cite{RossStuffing}. These samples were studied utilizing the SEQUOIA direct geometry time-of-flight spectrometer~\cite{Sequoia}, which is located at the Spallation Neutron Source at Oak Ridge National Laboratory. In an INS experiment the CEF excitations are manifested as dispersion-less features with the strongest scattering intensity expected at low $|Q|$ positions as a result of their magnetic origin. SEQUOIA is the ideal instrument for the investigation of the CEFs since it provides low $|Q|$ coverage and a large, dynamic ($|Q|$,E) range.

12g of each of the stoichiometric and stuffed powder samples were loaded in an aluminium flat plate with dimensions $\mathrm{50 mm \times 50 mm \times 1mm}$ and sealed with indium in helium atmosphere. An empty can with the same dimensions was loaded with the two samples on a three sample changer in a closed-cycle refrigerator. Measurements have been performed over a range of temperatures from $T$ = 5 K to 300 K, and utilizing neutrons with incident energy $\mathrm{E_i = 150 meV}$ giving an elastic energy resolution of $\pm$2.8 meV. The energy resolution improves with increasing energy transfer and is $\sim$1.4meV and $\sim$1meV for energy transfers of 80 and 115meV, respectively. The corresponding chopper settings selected were $\mathrm{T_0 = 120 Hz}$ and $\mathrm{FC_1 = 600 Hz}$. Similar measurements were conducted on the empty can for use as a background measurement.

\section{Inelastic Neutron Scattering from Crystal Field Excitations in $\bf{\rm Yb_2Ti_2O_7}$}

The INS spectra taken on both the stoichiometric and stuffed powder samples are shown in Fig.~3 for energies up to $\mathrm{E_i = 150 meV}$ at $T$ = 5 K. An empty can data set, taken as background at the same temperature, has been subtracted from both the sample data sets. As seen in Fig. 3, we observe a set of dispersion-less excitations arising from both the CEF transitions, as well as optical phonons. Below 60 meV, the intensity associated with the dispersion-less excitations increases with increasing $|$Q$|$, characteristic of inelastic scattering from phonons. In contrast, the excitations at 76.7 meV, 81.8 meV and 116.2 meV, highlighted with blue, grey and red arrows, respectively, in Fig. 3, show inelastic scattering which increases with decreasing $|$Q$|$, consistent with magnetic scattering. We therefore ascribe these three dispersion-less excitations with CEF transitions from the GS doublet.

\begin{figure}[h]
\includegraphics[width=8.6cm]{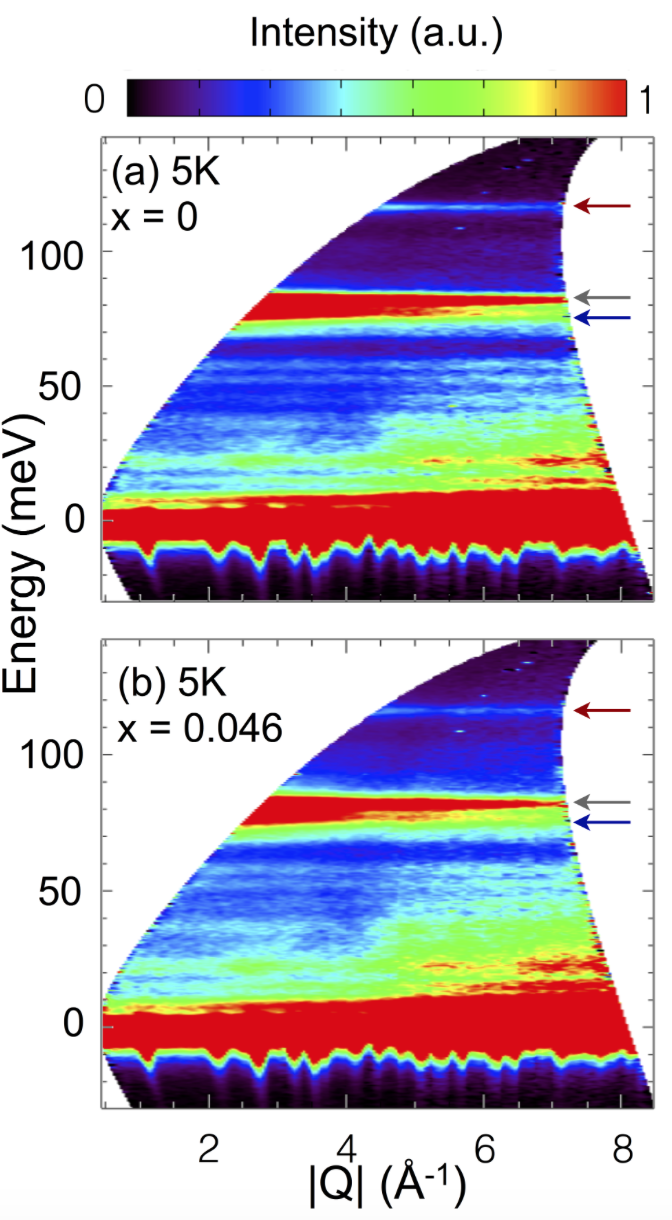}
\caption{\label{Emaps}\emph{Inelastic neutron scattering spectra--} $S(|Q|,\hbar\omega)$ obtained for the stoichiometric powder and the stuffed powder samples at $T$ = 5 K are shown in panels (a) and (b) respectively, with the corresponding $T$ = 5 K empty can subtracted from each data set. The three horizontal arrows in blue, gray and red highlight the three crystal field excitations which are found at 76.7 meV, 81.8 meV and 116.2 meV, respectively.}
\end{figure}

$\rm Yb^{3+}$ possesses thirteen electrons in its almost filled $4f$ shell and, as a consequence of Kramers' theorem, its eight-fold degenerate CEF levels can be maximally split into four doublets.  We associate the three 
magnetic excitations observed at $T$ = 5 K in both the stoichiometric and the stuffed powder samples shown in Fig.~3, with the transitions between the CEF GS doublet and the CEF excited 
state doublets, as indicated in the inset of Fig.~4. These three transitions between CEF doublets account for all the CEF states within this $J$ = 7/2 multiplet appropriate to Yb$^{3+}$.  As will be described in further detail, the CEF transitions observed at low temperatures in the stoichiometric sample are sharper in energy than those in the presence of light stuffing. We shall restrict our quantitative analysis of the CEF spectra to the case of the stoichiometric powder sample. Therefore we use the x = 0 data set in Fig.~3 a), and perform a $|Q|$ integrated cut ($|Q|$ = [4.5,5.25]\AA$^{-1}$), yielding the neutron scattering intensity as a function of energy. This data set is shown in the main panel of Fig.~4, wherein the intensity at the peak of the CEF transition at 81.8 meV has been normalized to unity. The relative intensities of the CEF transitions at 76.7 meV, 81.8 meV and 116.2 meV, as well as the energy of these transitions from the GS, constrain the CEF Hamiltonian.

The cut shown in Fig.~4 has been fitted to a model for the inelastic scattering~\cite{Squires} arising from dipole allowed transitions between the CEF GS doublet and the three excited states, using Eq. 3. The starting parameters in the CEF Hamiltonian for Yb$^{3+}$ were those determined by Bertin \textit{et al.}~\cite{Bertin2012} for Yb$_2$Ti$_2$O$_7$ within the point charge approximation.  This calculation yielded CEF transitions at $\sim$60 meV, $\sim$70 meV and $\sim$90 meV, in contrast to those determined experimentally in Fig.~3. These starting parameters were then refined and a best fit (shown in solid red) to the $|Q|$ integrated cut in Fig.~4 was obtained. All three CEF transitions were fit using the same resolution determined energy width.  In addition, relatively weak Lorenzian lineshapes near 56 meV, 70 meV and 100 meV phenomenologically describe the inelastic scattering from the phonons which are in near proximity to the CEFs, and improved the fit when included. As can be seen, the overall description of the INS data from the stoichiometric sample in Fig. 4 is very good.

\begin{figure}[h]
\includegraphics[width=8.6cm]{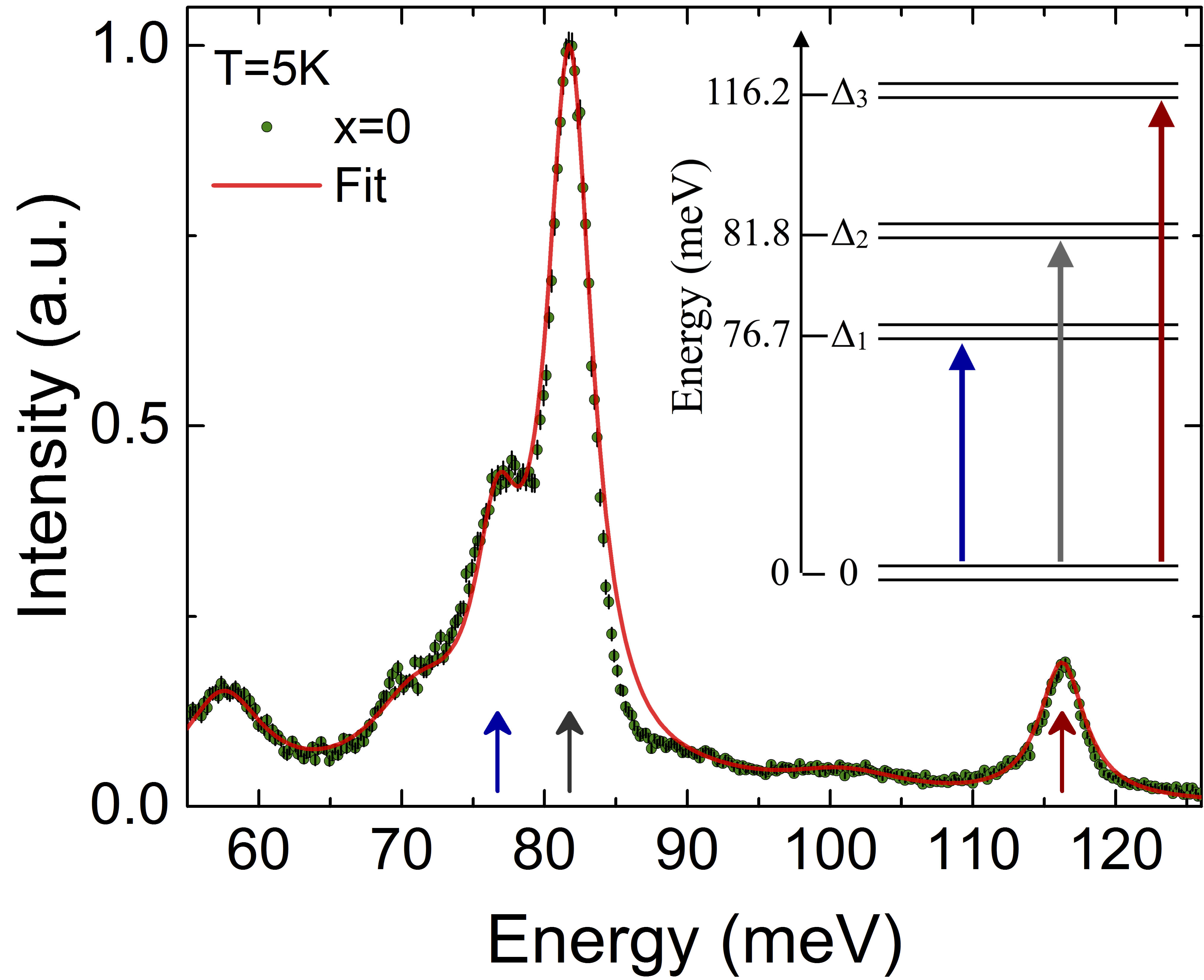}
\caption{\emph{$|Q|$ integrated cut for the stoichiometric powder at $T$ = 5 K. --}A $|Q|$ integrated cut ($|Q|$ = [4.5,5.25]\AA$^{-1}$) obtained from the INS spectra in Fig.3(a) for the stoichiometric powder at $T$ = 5 K is shown. The arrows indicate the corresponding positions of the CEFs. The inset outlines the corresponding CEF transitions from the GS doublet.}
\end{figure}

The values which were obtained for the CEF Hamiltonian parameters from the fit relevant to Yb$^{3+}$ at the A-site are given in Table~\ref{tab: CFParams} and the resulting energy eigenvalues and eigenvectors are given in Table ~\ref{tab: Eigen}. The GS doublet for Yb$^{3+}$ is comprised primarily of m$_J$ = $\pm$ 1/2 and the corresponding low temperature anisotropic g-tensor components are given by $g_\perp = 3.62$ $\pm$ 0.15 and $g_z = 1.85$ $\pm$ 0.10, where \textit{z} corresponds to the local [111] axis.  The error bars on the g values were estimated by exploring the sensitivity of the fitting procedure on the low temperature stoichiometric powder data, to different quantitative descriptions of the background phonons. Note that this determination of the g-tensor is performed at zero magnetic field, in contrast to, for example, the analysis of spin wave data in high magnetic fields~\cite{RossPRX}.  The g-tensor is consistent with previous estimates for Yb$^{3+}$ in Yb$_2$Ti$_2$O$_7$~\cite{Cao2009,Hodges2001,Malkin2004}, and the dominant m$_J$ = $\pm$ 1/2 character of the GS validates the effective S = 1/2 quantum description for the Yb$^{3+}$ moment. In order of ascending energy, the excited state doublets correspond primarily to m$_J$ = $\pm$ 7/2, m$_J$ = $\pm$ 3/2, and $m_J$ = $\pm$ 5/2.

\begin{table} 
\begin{tabular}{|c|c|c|c|}
\hline\hline
$B_n^m$ (meV) & Calculated & Fitted & Ratio  \\
\hline
$B_2^0$ & 1.270 & 1.135 & 0.894 \\
$B_4^0$ & $-0.0372$& $-0.0615$ & 1.653 \\
$B_4^3$ & $0.275$ & $0.315$ & 1.145 \\
$B_6^0$ & $0.00025$ & $0.0011$ & 4.4 \\
$B_6^3$ & $0.0023$ & $0.037$& 16.087 \\
$B_6^6$ & $0.0024$ & $0.005$ & 2.083 \\
\hline\hline
\end{tabular}
\caption{
\label{tab: CFParams}
A comparison of the calculated crystal field parameters ($B_n^m$) with those obtained by fitting INS data from the stoichiometric powder at T = 5 K.}
\end{table}

\begin{table*} 
\begin{tabular}{|L|RRRRRRRR|}
\hline\hline
E(meV) & $\Ket{-\frac{7}{2}}$ & $\Ket{-\frac{5}{2}}$ & $\Ket{-\frac{3}{2}}$ & $\Ket{-\frac{1}{2}}$ & $\Ket{\frac{1}{2}}$ & $\Ket{\frac{3}{2}}$ & $\Ket{\frac{5}{2}}$ & $\Ket{\frac{7}{2}}$ \\
\hline
0 & 0 & 0.0866 & 0 & 0 & -0.9283 & 0 & 0 & 0.3616 \\
\hline
0 & -0.3616 & 0 & 0 & -0.9283 & 0 & 0 & -0.0866 & 0 \\
\hline
76.706 & 0.9136 & 0 & 0 & -0.3343 & 0 & 0 & -0.2313 & 0 \\
\hline
76.706 & 0 & -0.2313 & 0 & 0 & 0.3343 & 0 & 0 & 0.9136 \\
\hline
81.764 & 0 & 0 & -1 & 0 & 0 & 0 & 0 & 0 \\
\hline
81.764 & 0 & 0 & 0 & 0 & 0 & -1 & 0 & 0 \\
\hline
116.23 & 0 & 0.9690 & 0 & 0 & 0.1627 & 0 & 0 & 0.1858 \\
\hline
116.23 & 0.1858 & 0 & 0 & -0.1627 & 0 & 0 & 0.9690 & 0 \\
\hline\hline
\end{tabular}
\caption{
\label{tab: Eigen}
\emph{The CEF eigenvalues and eigenvectors for Yb$^{3+}$ at the A-site of $\mathrm{Yb_2Ti_2O_7}$.} 
The first column displays the CEF eigenvalues of the system, while the corresponding eigenvectors 
are given in each row in terms of the $m_J$ basis.}
\end{table*}

\subsection{Temperature and weak stuffing dependence of the A-site CEF transitions}

Inelastic neutron scattering measurements have also been carried out on both the stoichiometric and stuffed powder samples as a function of temperature, using $\mathrm{E_i = 150 meV} $ neutrons. Energy scans of the $|Q|$ = [4.5,5.25]\AA$^{-1}$ integrated inelastic scattering for the stoichiometric powder sample are shown as a function of temperature in Fig. 5. The $T$ = 5 K data set is the same x = 0 data set shown in Fig. 4, and the peak intensity associated with the 81.8 meV CEF transition at $T$ = 5 K has been normalized to unity.

\begin{figure}[h]
\includegraphics[width=8.6cm]{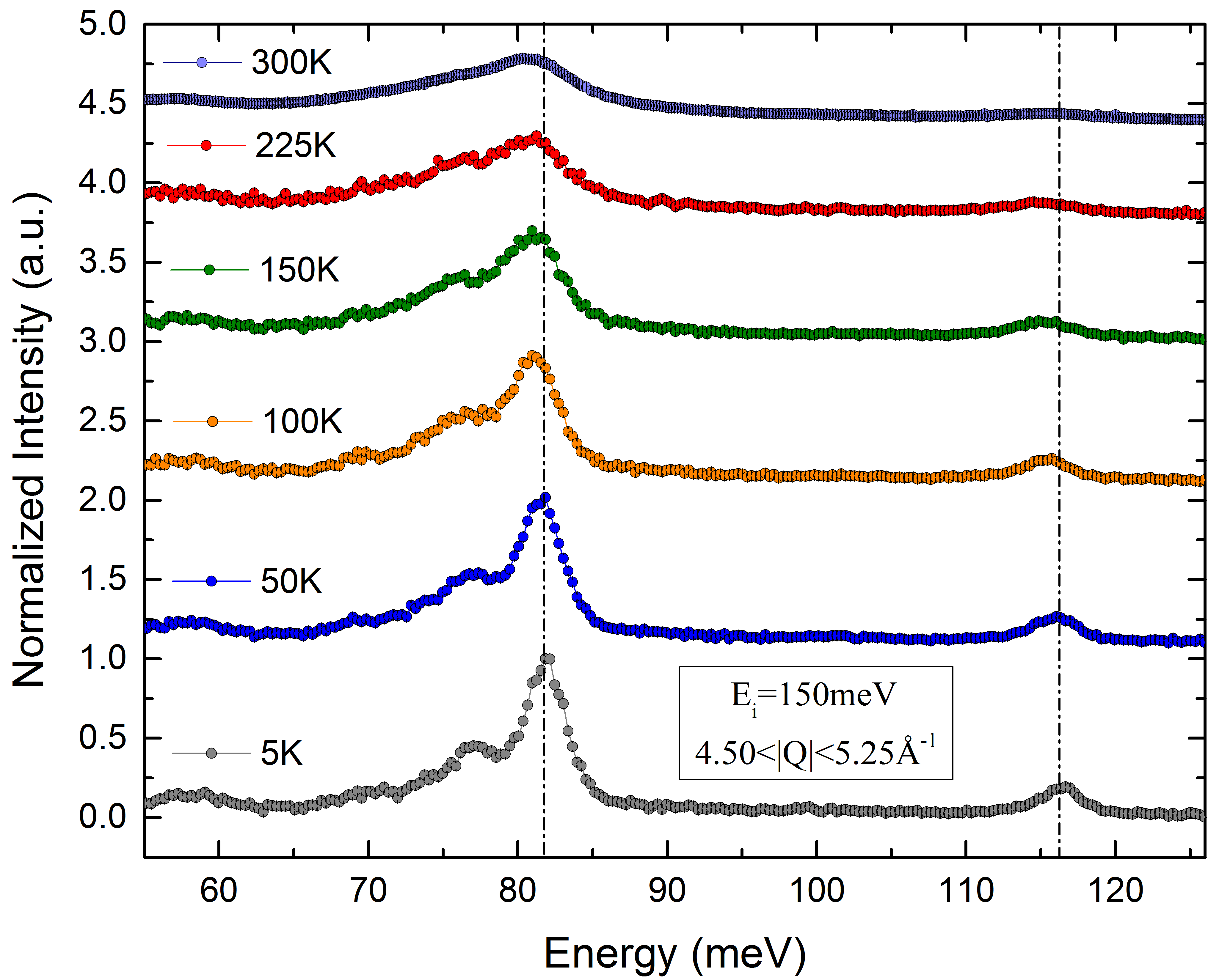}
\caption{\label{AllTempCEF} \emph{Energy cuts of the $|Q|$ = [4.5,5.25]\AA$^{-1}$ integrated inelastic scattering for the stoichiometric powder sample as a function of temperature. --} Energy cuts were taken with an appropriate background subtraction and all energy cuts for temperature above $T$ = 5 K has been vertically translated for clarity.}
\end{figure}

We note that the maximum temperature employed in these measurements, 300 K, corresponds $\approx$ 27 meV and consequently only the GS doublet of Yb$^{3+}$ is substantially occupied at any temperature. Qualitatively, the thermal fluctuations have three effects on the CEF neutron spectra: the CEF excitations broaden appreciably in energy; the maximum peak intensity diminishes; the energy of the CEF excitations softens slightly.  This latter effect is somewhat subtle, but it can be seen in Fig. 5 by drawing a fiducial dashed line positioned at the centre of the CEF transitions at $T$ = 5 K.

The energy width of the CEF excitations can be quantitatively examined by fitting the data sets as in Fig. 4, but now with a damped harmonic oscillator (DHO) lineshape for the three CEF transitions, each with 
the same temperature dependent energy width. At the energy transfers and temperatures of interest, the DHO can be approximated by a single Lorentzian for each mode.The form of this line-shape is given by:

\begin{eqnarray}
L(E) = \frac{1}{\pi} \frac{(\frac{\Gamma_{\text{obs}}}{2})}{(E-\Delta E)^{2}+(\frac{\Gamma_{\text{obs}}}{2})^{2}},
\end{eqnarray}
which is a Lorentzian function of energy with width $\Gamma_{\text{obs}}$ and centered on the energy of the CEF transition $\Delta E$. The falloff of the maximum peak intensity as a function of increasing temperature along with the concomitant broadening in the energy widths of the transitions ensures that the integrated spectral weight of this inelastic scattering is almost temperature independent, consistent with the CEF transitions being from the 
GS doublet of Yb$^{3+}$ at all temperatures considered here.  The common energy width or inverse lifetime of the three CEF excitations extracted from this analysis is plotted as a function of temperature in Fig. 6, where we removed the resolution contribution to the widths using the following relation :
\begin{eqnarray}
{\mathrm{\Gamma^2_{intrinsic}(T)} = {\Gamma^2_{\text{obs}}(T)} - {\Gamma^2_{\text{res}}}}.
\end{eqnarray}

Figure 6 then shows pronounced growth in the intrinsic energy width of the CEF transitions from $\sim$ 3.5 meV at $T$=5 K to $\sim$ 9 meV at $T$=300 K.  This growth mirrors the temperature dependence of the Yb$^{3+}$ mean squared displacements (MSDs) in stoichiometric Yb$_2$Ti$_2$O$_7$ as determined by powder neutron diffraction \cite{RossStuffing}, and this is also reproduced in Fig. 6 for direct comparison. It is worth noting that we plot one of the anisotropic components of the atomic displacement parameter, U$_{11}$, which represents the anisotropic MSD of the Yb$^{3+}$ ions from their average positions. Although, for the specific case of the Yb$^{3+}$ ion in the stoichiometric powder, it can be shown that U$_{11}$ is equal to the other components of the atomic displacement, so that the MSD is effectively isotropic. The inset to Fig.~6 shows the slight softening of the CEF energies with increasing temperature. The softening is most pronounced for $T$ $<$ 125 K, and then it flattens out at higher temperature. The significance of 125 K = 11.3 meV is not completely clear, although it could be related to the temperature scale associated with the top of the acoustic phonon band in Yb$_{2}$Ti$_{2}$O$_{7}$, which is typically 10 - 15 meV for transition metal oxides. This effect is relatively small with a maximum observed softening of $\sim$ 0.8 meV at $T$ = 300 K compared with T=5 K.

\begin{figure}[h]
\includegraphics[width=8.6cm]{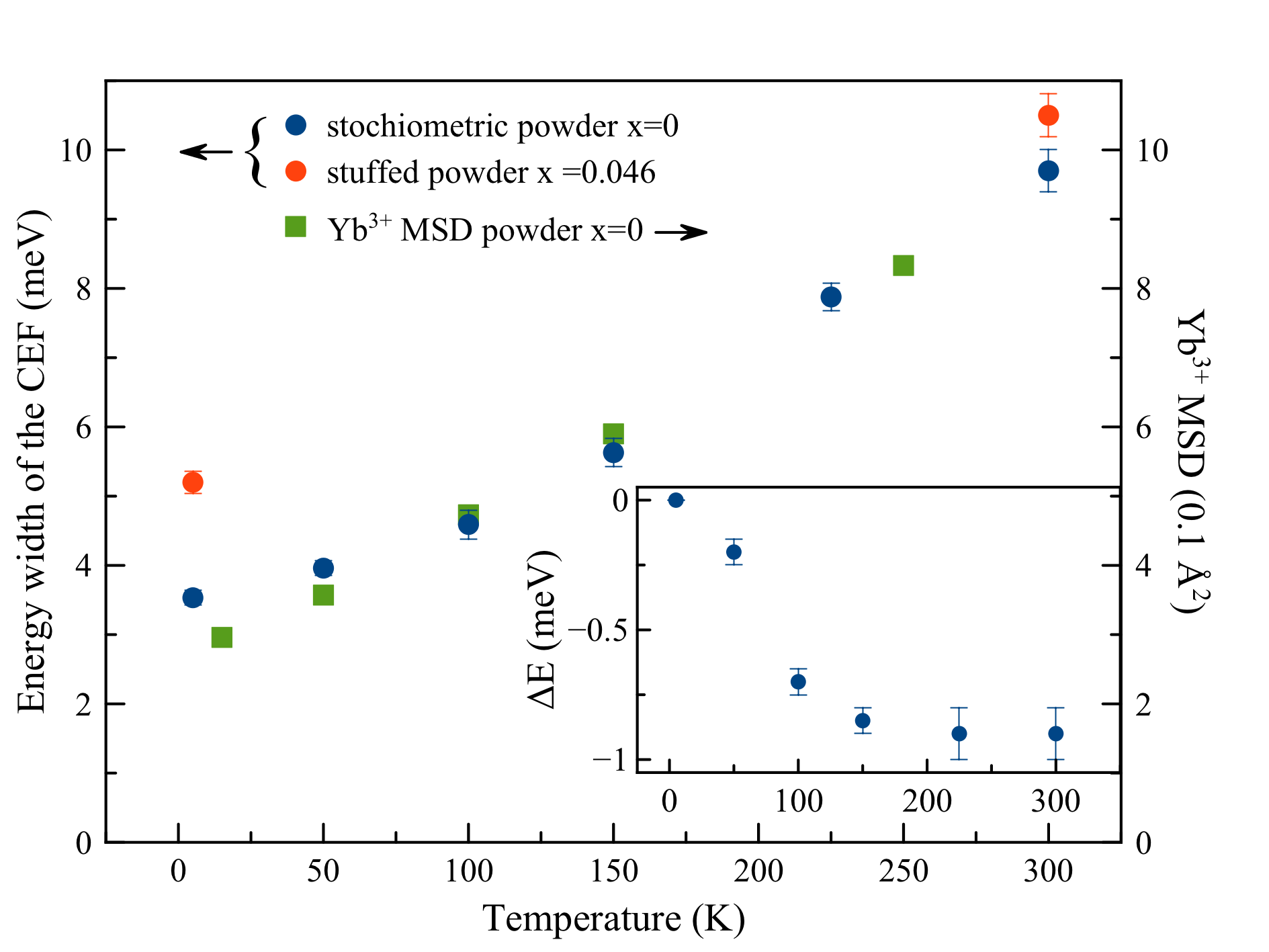}
\caption{\label{LifetimeTempCEF} \emph{Energy width and shift of the CEF transitions as a function of temperature--} The intrinsic energy width of the CEF transitions as extracted using Eq. 5. The intrinsic energy width corresponding to the stuffed powder is higher at both 5 K and 300 K in comparison to the stoichiometric sample. The green dots shows the mean square displacement (MSD) of the Yb$^{3+}$ taken from ref \cite{RossStuffing} and scaled in such a way that the MSDs and the energy widths of the CEF excitations in the stoichiometric powder have the same value at T=100K.  The inset shows the shift in energy of the CEF transitions as a function of temperature, relative to that at $T$=5 K.}
\end{figure}

The finite intrinsic energy widths of the CEF excitations in stoichiometric powder must originate from dynamics in the lattice not captured by the static structure.  The strong resemblance of the temperature dependence of the Yb$^{3+}$ MSDs to that of the CEF energy widths suggests that zero point fluctuations determine the low temperature energy widths, and the thermal population of phonons give rise to the larger widths at finite temperature.   The ionic displacements associated with both zero point fluctuations and the phonons will distort the local environment at the Yb$^{3+}$ site, giving rise to a distribution of CEF transitions energies.  Displacements of harmonic phonons time-average to zero, consequently the distribution of CEF transitions in the presence of the phonons is approximately centred on the zero temperature CEF transitions, and the transitions remain well defined at all temperatures.

Figure 7 shows a direct comparison of the $T$ = 5 K, $|Q|$ = [4.50,5.25] \AA$^{-1}$ integrated inelastic scattering for the stoichiometric and stuffed samples. This comparison clearly shows the CEF excitations in the lightly-stuffed sample have a considerably larger energy width even at low temperatures. The INS from the stuffed powder was also fit to the same DHO line-shape as was fit to the stoichiometric data, and the corresponding intrinsic energy widths at $T$ = 5 K and 300 K are also plotted in Fig. 6. The intrinsic energy widths of the CEF excitations in the stoichiometric and lightly stuffed samples differ by only $\sim$ 5 $\%$ at $T$ = 300 K; this is not surprising as the intrinsic energy widths in either sample are expected to be dominated by thermal fluctuations. However, at $T$ = 5 K, the intrinsic energy width of the stuffed sample is $\sim$ 5.2 meV - approximately that displayed by the stoichiometric sample at $\sim$ 125 K. 

\begin{figure}[h]
\includegraphics[width=8.6cm]{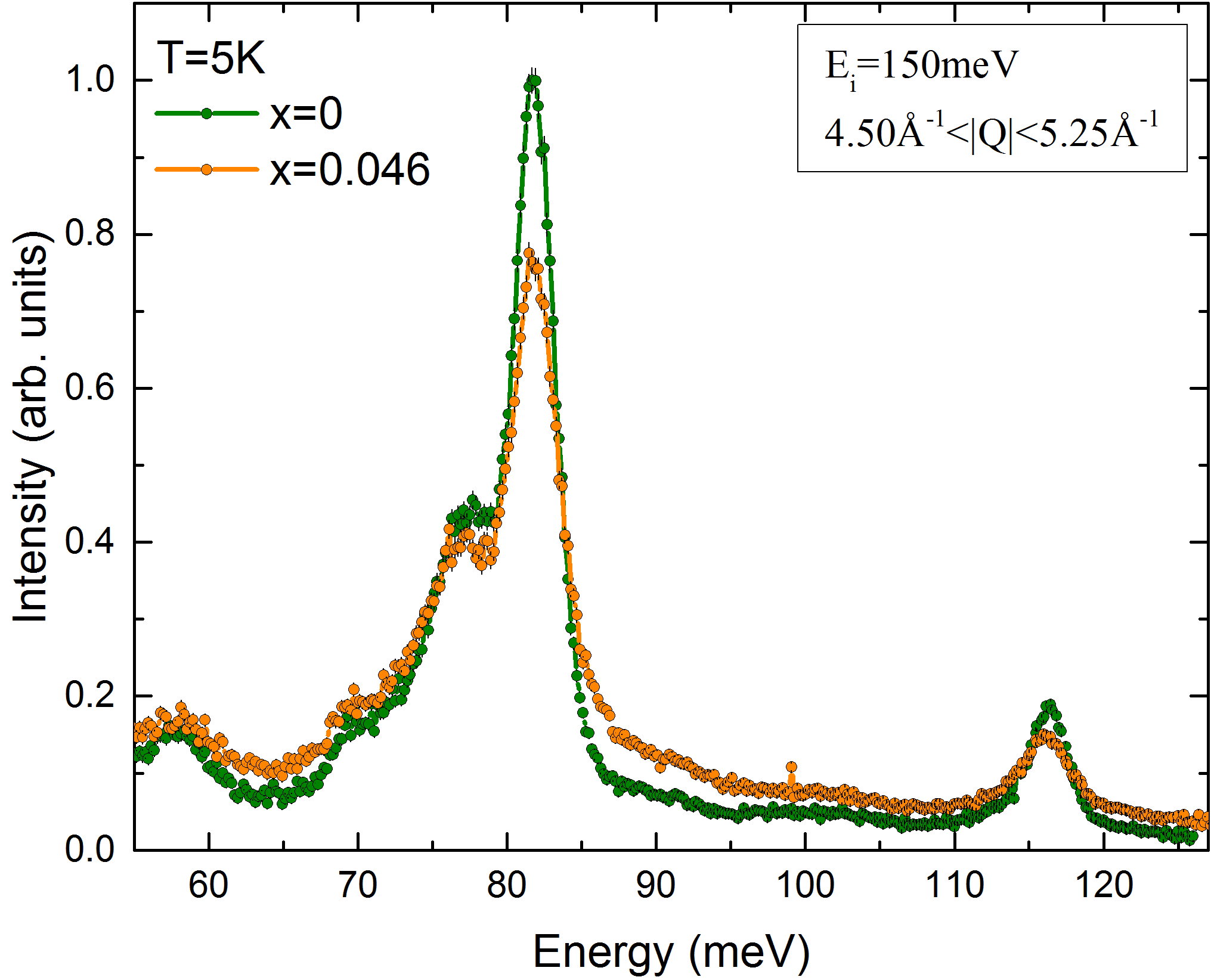}
\caption{\label{SampleDepCEF} \emph{A direct comparison between the INS from CEF transitions in the stuffed powder (in orange) and the stoichiometric powder (in green) $\rm Yb_2Ti_2O_7$ samples. --} Energy cuts taken at T = 5 K within a range $|Q|$ = [4.50,5.25]\AA$^{-1}$ for the stuffed powder (orange) and stoichiometric powder (green) samples.}
\end{figure}

The finite energy widths of the CEF excitations at low temperature in the lightly stuffed sample demonstrate that weak stuffing must induce a relatively large ``volume of influence" about each ``stuffed" Yb ion, that is, about each Yb ion residing on a B-site. The corresponding distortion field about each stuffed site gives rise to a distribution of A-site Yb$^{3+}$ environment, similar to those associated with phonons at finite temperature. Indeed, the stuffed B-site Yb ions would also give rise to CEF spectra completely distinct from those of Yb$^{3+}$ in A-site environments. However the concentration of such stuffed B-site Yb ions is known to be small, $\approx$ 2.3 $\%$, in our lightly stuffed sample and we were unable 
to find convincing evidence for such excitations in this study.

\section{Crystal Field Calculations from First Principles Within the Point Charge Approximation}

As discussed in Sec.~\ref{sec: CEF1}, the CEF Hamiltonian parameters can be fitted using INS data. In principle, these parameters can also be calculated from first principles within a point charge model following the procedure outlined in Refs.~\cite{Hutchings,Walter1984,Freeman1962}The CEF interactions are treated as a perturbation to the spin-orbit coupling and the resulting eigenfunctions of the CEF Hamiltonian are expressed as a linear combination of the $|J, m_J\rangle$ states within the $^2F_{\frac{7}{2}}$ manifold multiplets. 

We have calculated the CEF Hamiltonian parameters in this manner and the ratio between these ab initio parameters and those extracted from fit of the INS data are shown in Table~\ref{tab: CFParams}. Note that the ionic positions were taken from the crystallographic refinement of stoichiometric Yb$_2$Ti$_2$O$_7$ \cite{RossStuffing}.  As can be seen, apart from the $B^0_6$ and $B^3_6$ terms which are relatively small, all the CEF parameters agree well with those extracted from fitting the INS data. This result is remarkable given the simplicity of the point charge approximation, and it gives us confidence that we can make reasonable predictions on the strength of the CEF Hamiltonian parameters determined in this way.

The CEF eigenvectors and eigenvalues determined from the fit to the INS in the stoichiometric powder at $T$ = 5 K is shown in Table~\ref{tab: Eigen}. Tables III and IV show the CEF eigenvectors and eigenvalues for Yb ions in two different impurity sites: one ``stuffed" on the B-site and one on the A-site but in the presence of one O(1) oxygen vacancy. Both of these impurities are expected to be present at some small concentration in the lightly stuffed sample. In principle O(2) vacancies can also occur, but these are higher energy defects than O(1) vacancies ~\cite{Sala2014}.  These CEF eigenvectors and eigenvalues have been calculated within the point charge approximation and then corrected with the same ratio of $B^m_n$ parameters as were determined from the comparison between the INS fitted values and the calculated values in stoichiometric Yb$_2$Ti$_2$O$_7$ at low temperatures. 

The differences between the fitted CEF parameters and the calculated ones originate mainly from the overlap of the 4f orbital of the rare earth with the 2p orbitals of the ligands. The corresponding effects on the $B^m_n$ terms are difficult to estimate from first principles and are not considered here.  Instead we calculate the CEF parameters for the defective environments, and then correct these with the same $B^m_n$ ratios which were determined in the treatment of CEFs for the stoichiometric sample.  Finally, we verified that this correction did not dramatically affect the properties of the system; that is that the CEF properties within the defective environments did not qualitatively depend on this correction.

\subsection{Crystal Field Calculation for Yb$^{3+}$ at the A-site in an Oxygen Depleted Environment}

We first consider an A-site Yb$^{3+}$ in the presence of a single O(1) oxygen vacancy. This depleted environment is expected to break the symmetries of the CEF Hamiltonian. However if Prather's convention is satisfied, only the inversion operation is lost and the number of CEF parameters in our Hamiltonian is unchanged from the stoichiometric case. Note that in general, a broken symmetry can dramatically affect the CEF Hamiltonian, and it may be necessary to add terms to the CEF Hamiltonian to better approximate the defective Coulomb potential.

\begin{table*} 
\begin{tabular}{|L|RRRRRRRR|}
\hline\hline
E(meV) & $\Ket{-\frac{7}{2}}$ & $\Ket{-\frac{5}{2}}$ & $\Ket{-\frac{3}{2}}$ & $\Ket{-\frac{1}{2}}$ & $\Ket{\frac{1}{2}}$ & $\Ket{\frac{3}{2}}$ & $\Ket{\frac{5}{2}}$ & $\Ket{\frac{7}{2}}$ \\
\hline
0 & -0.9933 & 0 & 0 & -0.1136 & 0 & 0 & 0.02 & 0 \\
\hline
0 & 0 & -0.02 & 0 & 0 & -0.1136 & 0 & 0 & 0.9933 \\
\hline
163.789 & 0 & 0.9844 & 0 & 0.0010 & -0.1756 & 0 & 0.0056 & 0 \\
\hline
163.789 & 0 & 0.0056 & 0 & -0.1756 & -0.0010 & 0 & -0.9844 & 0 \\
\hline
230.125 & 0 & 0 & 0 & 0 & 0 & -1 & 0 & 0 \\
\hline
230.125 & 0 & 0 & 1 & 0 & 0 & 0 & 0 & 0 \\
\hline
238.306 & 0.1153 & -0.0061 & 0 & -0.9773 & -0.0341 & 0 & 0.1743 & -0.0040 \\
\hline
238.306 & -0.0040 & -0.1743 & 0 & 0.0341 & -0.9773 & 0 & -0.0061 & -0.1153 \\
\hline\hline
\end{tabular}
\caption{
\label{tab: Eigen2}
\emph{The CEF eigenvalues and eigenvectors calculated for Yb$^{3+}$ in a depleted oxygen environment at the A-site in $\mathrm{Yb_2Ti_2O_7}$.} 
The first column displays the CEF eigenvalues of the system, while the corresponding eigenvectors 
are given in each row in terms of the $m_J$ basis. The ratio correction was applied in order to arrive at the final CEF parameters.}
\end{table*}

The eigenvalues and eigenvectors calculated for A-site Yb$^{3+}$ in the presence of a single O(1) oxygen vacancy are shown in Table~\ref{tab: Eigen2}. As previously discussed, we have performed the point charge calculation and then scaled these results by the $B^m_n$ ratios taken from the fitted and calculated CEF Hamiltonian terms determined for the stoichiometric powder.

Kramers' degeneracy still protects the Yb$^{3+}$ ion in this defective environment and all the CEF eigenvectors appear as doublets. However, the gap between the GS and the first excited state is much larger than in the stoichiometric case. Moreover an examination of the eigenvectors in Table~\ref{tab: Eigen2} shows that the GS is now a pure linear combination of $m_J$ = $\pm 7/2$   states. These GS eigenvectors are similar to what is found in stoichiometric \DTO~crystals, where the GS is comprised of a linear combination of the maximal $m_J$  states, pure $m_J$=$\pm 15/2$.

\begin{figure}[h]
\includegraphics[width=8.5cm]{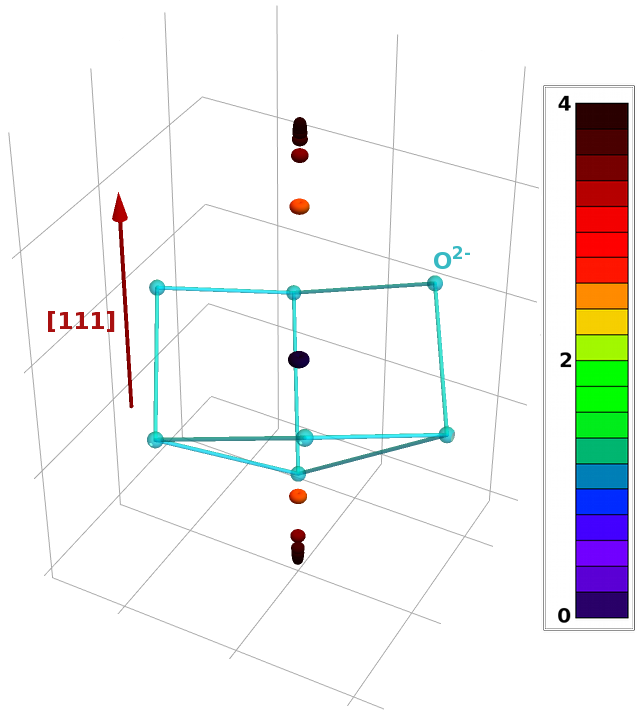}
\label{fig: Vacancy}
\caption{\emph{Magnetization of the A-site $\rm Yb^{3+}$ ion in presence of an O(1) oxygen vacancy. --} A coloured sphere represents the tip of the magnetization vector centred at the $\rm Yb^{3+}$ ion in response to an applied external field of H = 1 T which rotates within the plane normal to [111].  Many values of the magnetization are shown simultaneously for applied fields uniformly distributed on the unit sphere. For the same [111] component of magnetization, the centre of the spheres are sufficiently close to each other, that the final impression is that of rings.  The tightness of the rings around the [111] axis implies stronger Ising-like behaviour of the Yb$^{3+}$ moment, in this case with a calculated magnetic moment of $3.953\mu_B$ along the local [111] direction.}
\end{figure}

With the CEF eigenfunctions in hand, we can calculate the local magnetization of the A-site Yb$^{3+}$ in the presence of a single O(1) oxygen vacancy.  This is shown in Fig. 8 where the magnetization anisotropy for $\rm Yb^{3+}$ moment within the defective scalenohedron is displayed.  In both Figs. 8 and 9, coloured spheres represent the tips of magnetization vector centred at the $\rm Yb^{3+}$ ion, in response to a given applied external field of H = 1 T with a particular [111]-component and which precesses around the plane normal to [111].  Many values of the magnetization are shown simultaneously for applied fields uniformly distributed on the unit sphere. The colour scale is chosen in order to match the strength of the magnetic moment from $0\mu_B$ (dark blue) to $4\mu_B$ (dark red). For the same [111] component of magnetization, the centre of the spheres are sufficiently close to each other, that the final impression is that of rings (this is seen more clearly in Fig. 9).  The tightness of the rings about the [111] axis implies stronger Ising-like behaviour of the Yb$^{3+}$ moment.  For the case of the A-site Yb$^{3+}$ in the presence of a single O(1) oxygen vacancy in Fig. 8, the calculated magnetic moment along the local [111] direction is $3.953\mu_B$, close to the full moment of $4 \mu_B$.  This suggests negligible precession around this easy axis and thus strong Ising-like behavior.  The local magnetization anisotropy for the A-site Yb$^{3+}$ in the presence of a single O(1) oxygen vacancy is dramatically different from the stoichiometric case, where it was planar or XY-like~\cite{Hodges2001,Malkin2004,Siddhartan1999}, and is similar to that displayed by the classical spin ice magnets 
\DTO and \HTO. This anisotropy is present even without the $B^m_n$ ratio corrections and thus it depends only on the depleted environment.

\subsection{Crystal Field Calculation for ``Stuffed" Yb$^{3+}$ at the B-site}

Neutron powder diffraction measurements on the stuffed powder samples, derived from crushed single crystals, show such Yb$_2$Ti$_2$O$_7$ samples to be lightly stuffed at the 2.3 $\%$ level.  We therefore calculate the CEF eigenvectors and eigenvalues for ``stuffed" Yb$^{3+}$ at the B-site.  This is an interesting case to consider since the local environment at the B-site is very different from that considered to this point for A-site Yb${3+}$. Nonetheless, the symmetry at the B-site is similar to the A-site, and we can simply rotate the trigonal anti-prism cage around  the B-site so that the local $[111]$ direction is aligned along $\hat{z}$ and the $C_2$ axis along $\hat{y}$. Therefore the form of the B-site Yb$^{3+}$ CEF Hamiltonian is identical to Eq.~\ref{eq: HCEF}.  The calculated CEF eigenvectors and eigenvalues for B-site Yb$^{3+}$ are displayed in Table~\ref{tab: Eigen3}. Once again we have performed the point charge calculation and then scaled these results by the $B^m_n$ ratios determined for the stoichiometric powder.\\

\begin{table*} 
\begin{tabular}{|L|RRRRRRRR|}
\hline\hline
E(meV) & $\Ket{-\frac{7}{2}}$ & $\Ket{-\frac{5}{2}}$ & $\Ket{-\frac{3}{2}}$ & $\Ket{-\frac{1}{2}}$ & $\Ket{\frac{1}{2}}$ & $\Ket{\frac{3}{2}}$ & $\Ket{\frac{5}{2}}$ & $\Ket{\frac{7}{2}}$ \\
\hline
0 & 0 & -0.2733 & 0 & 0 & -0.2648 & 0 & 0 & 0.9248 \\
\hline
0 & 0.9248 & 0 & 0 & 0.2648 & 0 & 0 & -0.2733 & 0 \\
\hline
69.991 & 0 & -0.9558 & 0 & 0 & -0.0338 & 0 & 0 & -0.2921 \\
\hline
69.991 & -0.2921 & 0 & 0 & 0.0338 & 0 & 0 & -0.9558 & 0 \\
\hline
247.81 & 0 & 0 & 0.1238 & 0 & 0 & -0.9923 & 0 & 0 \\
\hline
247.81 & 0 & 0 & -0.9923 & 0 & 0 & -0.1238 & 0 & 0 \\
\hline
318.797 & 0.2439 & 0 & 0 & -0.9637 & 0 & 0 & -0.1086 & 0 \\
\hline
318.797 & 0 & 0.1086 & 0 & 0 & -0.9637 & 0 & 0 & -0.2439\\
\hline\hline
\end{tabular}
\caption{
\label{tab: Eigen3}
\emph{The CEF eigenvalues and eigenvectors calculated for $\mathrm{Yb_2Ti_2O_7}$ at B-site.} 
The first column displays the Crystal Field spectrum of the system, while the corresponding eigenvectors 
are given in each row in terms of $m_J$ basis. The ratio correction was applied in this case to the CEF parameters.}
\end{table*}

Kramers' degeneracy is again invoked and all CEF levels are again doublets, now with a gap of $105$ meV between the GS and the first excited state. The eigenvectors within the GS doublet are again almost pure m$_J$=$\pm 7/2$ spin states, and we expect the anisotropy to be Ising like. This is borne out by again applying a small rotating external magnetic field to probe the shape of the anisotropy of the GS magnetic moment as was done for  Yb$^{3+}$ at the A-site in an oxygen depleted environment. Following the previous convention, Fig. 9 shows spheres centred along the direction in which the magnetic moment points. The colour scale matches the magnetic moment size in units of $\mu_B$. In contrast with the previous case, the spheres are not located exactly along the local $[111]$ direction, thus we anticipate a small precession about the easy axis with the maximum value of the magnetic moment given by $\mu = 3.25\mu_B$.  This is similar to what was calculated for the defective scalenohedron associated with Yb$^{3+}$ at the A-site in an oxygen depleted environment, but the anisotropy is not as Ising-like as it was for that case.

\begin{figure}[h]
\includegraphics[width=8.5cm]{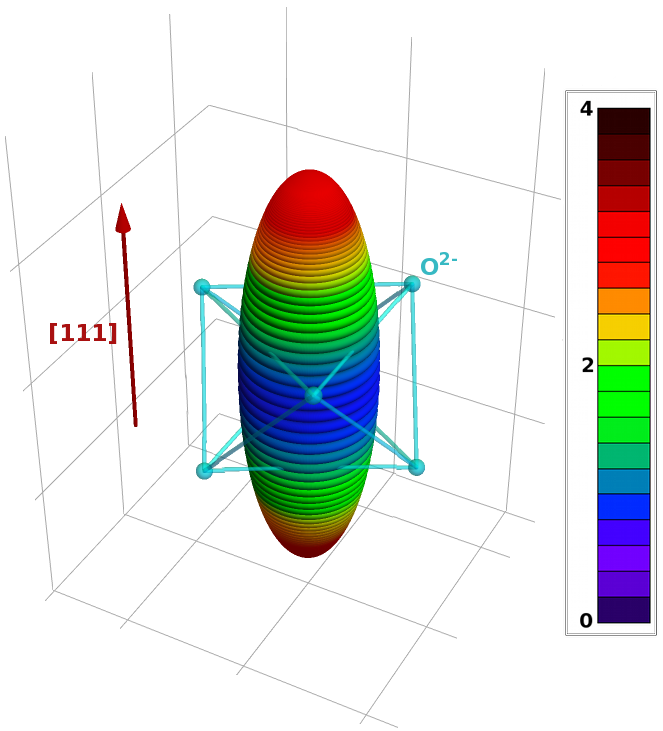}
\label{fig: StufDef}
\caption{\emph{Magnetization of the ``stuffed" $Yb^{3+}$ ion at the B-site of $\mathrm{Yb_2Ti_2O_7}$. --} Each sphere
represents the tip of the magnetization vector centred at the $\rm Yb^3+$ ion in response to an applied rotating external field of H = 1 T. Neighboring contiguous spheres form ring pattern. The ellipsoidal shape of the ring pattern suggests Ising-like behaviour of the rare earth moment along the local [111] direction, however the transverse extent of the ellipsoidal shape indicates a small precession around the Ising axis. The calculated magnetic moment is $3.25\mu_B$.}
\end{figure}

It is interesting that the influence of oxygen vacancies on A-site Yb$^{3+}$ ions and that of stuffed B-site Yb$^{3+}$ ions have similar effects on the CEF eigenvalues and eigenvectors. Stuffing Yb$^{3+}$ onto B-sites nominally occupied by Ti$^{4+}$ will require oxygen vacancies to preserve charge neutrality, and we expect both phenomena to occur at some level in single crystals grown from the melt. The symmetry of the defective scalenohedron, especially when two O(1) are missing, is reasonably similar and to the trigonal anti-prism associated with stuffed Yb$^{3+}$ at the B-site, so it is perhaps not so surprising that these environments produce similar effects on the CEF eigenvalues and eigenvectors of the magnetic ions. On the other hand the similarity of the CEFs associated with the defective sites may complicate the analysis of  experimental data from sufficiently defective non-stoichiometric Yb$_2$Ti$_2$O$_7$, as the measured spectrum may show overlap originating from at least two distinct defective environments.

\section{CONCLUSIONS}  

Neutron scattering measurements on the stoichiometric and stuffed powder samples of the quantum spin ice candidate system Yb$_2$Ti$_2$O$_7$ have been carried out to probe the CEF eigenvalues and eigenvectors associated with the $J$ = 7/2 Yb$^{3+}$ ion in these environments. Analysis of the INS from the stoichiometric powder show the GS doublet to correspond primarily to m$_J$ = $\pm$ 1/2, with anisotropic g-tensor components given by $g_\perp = 3.62$ $\pm$ 0.15 and $g_z = 1.85$ $\pm$ 0.10, which therefore imply local XY anisotropy as expected. The eigenvalues and eigenvectors of the three excited state CEF doublets were all identified above a lowest CEF gap of $\sim$ 76.7 meV at $T$ = 5 K.  The energy widths, or inverse lifetimes of the CEF states broaden with increasing temperature from $\sim$ 3.5 meV at low temperatures up to the highest measured temperature, 300 K. This broadening mirrors the temperarure dependence of the Yb$^{3+}$ MSDs, and we associate it with the effects of zero point motion and phonons on the CEF environment around the A-site Yb$^{3+}$ site. Interestingly, the corresponding measurements and analysis of the lightly stuffed powder sample (x = 0.046) of Yb$_2$Ti$_2$O$_7$ shows a similar set of CEF transitions, however the energy width or inverse lifetime of the transitions are intrinsically further broadened even at $T$ = 5 K. This effect is ascribed to a relatively large strain field, or ``volume-of-influence", associated with each stuffed B-site Yb$^{3+}$. 

We have also calculated the CEF eigenvectors and eigenvalues associated with Yb$^{3+}$ in two defective environments of relevance for non-stoichiometric Yb$_2$Ti$_2$O$_7$. These are for Yb$^{3+}$ at the A-site in an oxygen depleted environment, and for stuffed Yb$^{3+}$ at the B-site. These calculations were performed within the point charge approximation and make use of a comparison of the calculated and measured eigenvalues and eigenfunctions in the stoichiometric sample to benchmark these results.  Both of these defective environments give rise to related effects wherein the GS doublet is now primarily made up of m$_J$ = $\pm$ 7/2, and the local anisotropy of the GS eigenfunctions are Ising-like, although with stronger Ising anisotropy for the oxygen depleted environment than for case of the stuffed Yb$^{3+}$ at the B-site.

These results put the nature of the CEF eigenfunctions and eigenvalues associated with Yb$^{3+}$ in the quantum spin ice candidate system on a much firmer footing and indicate how the nature of the GS moment and anisotropy in this system can be sensitive to small amounts of disorder, as is known to occur in single crystal samples grown from the melt using floating zone techniques.

\begin{acknowledgments}
We would like to acknowledge helpful conversations with M. J. P. Gingras. We would also like to acknowledge T. E. Sherline for technical assistance with the measurements. The neutron scattering data were reduced using Mantid~\cite{Mantid} and analyzed using the DAVE software package~\cite{Dave}.  Research using ORNL's Spallation Neutron Source was sponsored by the Scientific User Facilities Division, Office of Basic Energy Sciences, U.S. Department of Energy. Work at McMaster University was funded by NSERC of Canada.
\end{acknowledgments}

\bibliography{bibYTO.bib}

\begin{thebibliography}{43}
\expandafter\ifx\csname natexlab\endcsname\relax\def\natexlab#1{#1}\fi
\expandafter\ifx\csname bibnamefont\endcsname\relax
  \def\bibnamefont#1{#1}\fi
\expandafter\ifx\csname bibfnamefont\endcsname\relax
  \def\bibfnamefont#1{#1}\fi
\expandafter\ifx\csname citenamefont\endcsname\relax
  \def\citenamefont#1{#1}\fi
\expandafter\ifx\csname url\endcsname\relax
  \def\url#1{\texttt{#1}}\fi
\expandafter\ifx\csname urlprefix\endcsname\relax\def\urlprefix{URL }\fi
\providecommand{\bibinfo}[2]{#2}
\providecommand{\eprint}[2][]{\url{#2}}

\bibitem[{\citenamefont{Lacroix et~al.}(2011)\citenamefont{Lacroix, Mendels,
  and Mila}}]{Lacroix}
\bibinfo{author}{\bibfnamefont{C.}~\bibnamefont{Lacroix}},
  \bibinfo{author}{\bibfnamefont{P.}~\bibnamefont{Mendels}}, \bibnamefont{and}
  \bibinfo{author}{\bibfnamefont{F.}~\bibnamefont{Mila}},
  \bibinfo{journal}{Springer Series in Solid-State Sciences
  (Springer,Heidelber)}  (\bibinfo{year}{2011}).

\bibitem[{\citenamefont{Subramanian et~al.}(1983)\citenamefont{Subramanian,
  Aravamudan, and Rao}}]{Subramanian}
\bibinfo{author}{\bibfnamefont{M.}~\bibnamefont{Subramanian}},
  \bibinfo{author}{\bibfnamefont{G.}~\bibnamefont{Aravamudan}},
  \bibnamefont{and} \bibinfo{author}{\bibfnamefont{G.~S.} \bibnamefont{Rao}},
  \bibinfo{journal}{Progress in Solid State Chemistry}
  \textbf{\bibinfo{volume}{15}}, \bibinfo{pages}{55 } (\bibinfo{year}{1983}),
  ISSN \bibinfo{issn}{0079-6786}.

\bibitem[{\citenamefont{Gardner et~al.}(2010)\citenamefont{Gardner, Gingras,
  and Greedan}}]{Greedan}
\bibinfo{author}{\bibfnamefont{J.~S.} \bibnamefont{Gardner}},
  \bibinfo{author}{\bibfnamefont{M.~J.~P.} \bibnamefont{Gingras}},
  \bibnamefont{and} \bibinfo{author}{\bibfnamefont{J.~E.}
  \bibnamefont{Greedan}}, \bibinfo{journal}{Rev. Mod. Phys.}
  \textbf{\bibinfo{volume}{82}}, \bibinfo{pages}{53} (\bibinfo{year}{2010}).

\bibitem[{\citenamefont{Bramwell and Gingras}(2001)}]{Bramwell2001}
\bibinfo{author}{\bibfnamefont{S.~T.} \bibnamefont{Bramwell}} \bibnamefont{and}
  \bibinfo{author}{\bibfnamefont{M.~J.~P.} \bibnamefont{Gingras}},
  \bibinfo{journal}{Science} \textbf{\bibinfo{volume}{294}},
  \bibinfo{pages}{1495} (\bibinfo{year}{2001}).

\bibitem[{\citenamefont{Castelnovo et~al.}(2008)\citenamefont{Castelnovo,
  Moessner, and Sondhi}}]{Castelnovo2008}
\bibinfo{author}{\bibfnamefont{C.}~\bibnamefont{Castelnovo}},
  \bibinfo{author}{\bibfnamefont{R.}~\bibnamefont{Moessner}}, \bibnamefont{and}
  \bibinfo{author}{\bibfnamefont{S.~L.} \bibnamefont{Sondhi}},
  \bibinfo{journal}{Nature} \textbf{\bibinfo{volume}{451}}, \bibinfo{pages}{42}
  (\bibinfo{year}{2008}).

\bibitem[{\citenamefont{den Hertog and Gingras}(2000)}]{DenHertog2000}
\bibinfo{author}{\bibfnamefont{B.~C.} \bibnamefont{den Hertog}}
  \bibnamefont{and} \bibinfo{author}{\bibfnamefont{M.~J.~P.}
  \bibnamefont{Gingras}}, \bibinfo{journal}{Phys. Rev. Lett.}
  \textbf{\bibinfo{volume}{84}}, \bibinfo{pages}{3430} (\bibinfo{year}{2000}).

\bibitem[{\citenamefont{Rosenkranz et~al.}(2000)\citenamefont{Rosenkranz,
  Ramirez, Hayashi, Cava, Siddharthan, and Shastry}}]{Rosenkranz1}
\bibinfo{author}{\bibfnamefont{S.}~\bibnamefont{Rosenkranz}},
  \bibinfo{author}{\bibfnamefont{A.~P.} \bibnamefont{Ramirez}},
  \bibinfo{author}{\bibfnamefont{A.}~\bibnamefont{Hayashi}},
  \bibinfo{author}{\bibfnamefont{R.}~\bibnamefont{Cava}},
  \bibinfo{author}{\bibfnamefont{R.}~\bibnamefont{Siddharthan}},
  \bibnamefont{and} \bibinfo{author}{\bibfnamefont{B.~S.}
  \bibnamefont{Shastry}}, \bibinfo{journal}{J.Appl.Phys}
  \textbf{\bibinfo{volume}{87}} (\bibinfo{year}{2000}).

\bibitem[{\citenamefont{Harris et~al.}(1997)\citenamefont{Harris, Bramwell,
  McMorrow, Zeiske, and Godfrey}}]{Harris1997}
\bibinfo{author}{\bibfnamefont{M.~J.} \bibnamefont{Harris}},
  \bibinfo{author}{\bibfnamefont{S.~T.} \bibnamefont{Bramwell}},
  \bibinfo{author}{\bibfnamefont{D.~F.} \bibnamefont{McMorrow}},
  \bibinfo{author}{\bibfnamefont{T.}~\bibnamefont{Zeiske}}, \bibnamefont{and}
  \bibinfo{author}{\bibfnamefont{K.~W.} \bibnamefont{Godfrey}},
  \bibinfo{journal}{Phys. Rev. Lett.} \textbf{\bibinfo{volume}{79}},
  \bibinfo{pages}{2554} (\bibinfo{year}{1997}).

\bibitem[{\citenamefont{Pauling}(1935)}]{Pauling}
\bibinfo{author}{\bibfnamefont{L.}~\bibnamefont{Pauling}},
  \bibinfo{journal}{Journal of the American Chemical Society}
  \textbf{\bibinfo{volume}{57}}, \bibinfo{pages}{2680} (\bibinfo{year}{1935}).

\bibitem[{\citenamefont{Yasui et~al.}(2003)\citenamefont{Yasui, Soda, Iikubo,
  Ito, Sato, Hamaguchi, Matsushita, Wada, Takeuchi, Aso et~al.}}]{Yasui2003}
\bibinfo{author}{\bibfnamefont{Y.}~\bibnamefont{Yasui}},
  \bibinfo{author}{\bibfnamefont{M.}~\bibnamefont{Soda}},
  \bibinfo{author}{\bibfnamefont{S.}~\bibnamefont{Iikubo}},
  \bibinfo{author}{\bibfnamefont{M.}~\bibnamefont{Ito}},
  \bibinfo{author}{\bibfnamefont{M.}~\bibnamefont{Sato}},
  \bibinfo{author}{\bibfnamefont{N.}~\bibnamefont{Hamaguchi}},
  \bibinfo{author}{\bibfnamefont{T.}~\bibnamefont{Matsushita}},
  \bibinfo{author}{\bibfnamefont{N.}~\bibnamefont{Wada}},
  \bibinfo{author}{\bibfnamefont{T.}~\bibnamefont{Takeuchi}},
  \bibinfo{author}{\bibfnamefont{N.}~\bibnamefont{Aso}}, \bibnamefont{et~al.},
  \bibinfo{journal}{Journal of the Physical Society of Japan}
  \textbf{\bibinfo{volume}{72}}, \bibinfo{pages}{3014} (\bibinfo{year}{2003}).

\bibitem[{\citenamefont{Chang et~al.}(2012)\citenamefont{Chang, Onoda, Su, Kao,
  Tsuei, Yasui, Kakurai, and Lees}}]{Chang2012}
\bibinfo{author}{\bibfnamefont{L.-J.} \bibnamefont{Chang}},
  \bibinfo{author}{\bibfnamefont{S.}~\bibnamefont{Onoda}},
  \bibinfo{author}{\bibfnamefont{Y.}~\bibnamefont{Su}},
  \bibinfo{author}{\bibfnamefont{Y.-J.} \bibnamefont{Kao}},
  \bibinfo{author}{\bibfnamefont{K.-D.} \bibnamefont{Tsuei}},
  \bibinfo{author}{\bibfnamefont{Y.}~\bibnamefont{Yasui}},
  \bibinfo{author}{\bibfnamefont{K.}~\bibnamefont{Kakurai}}, \bibnamefont{and}
  \bibinfo{author}{\bibfnamefont{M.~R.} \bibnamefont{Lees}},
  \bibinfo{journal}{Nat Commun} \textbf{\bibinfo{volume}{3}},
  \bibinfo{pages}{992} (\bibinfo{year}{2012}).

\bibitem[{\citenamefont{Hodges et~al.}(2002)\citenamefont{Hodges, Bonville,
  Forget, Yaouanc, Dalmas~de R\'eotier, Andr\'e, Rams, Kr\'olas, Ritter,
  Gubbens et~al.}}]{Hodges2002}
\bibinfo{author}{\bibfnamefont{J.~A.} \bibnamefont{Hodges}},
  \bibinfo{author}{\bibfnamefont{P.}~\bibnamefont{Bonville}},
  \bibinfo{author}{\bibfnamefont{A.}~\bibnamefont{Forget}},
  \bibinfo{author}{\bibfnamefont{A.}~\bibnamefont{Yaouanc}},
  \bibinfo{author}{\bibfnamefont{P.}~\bibnamefont{Dalmas~de R\'eotier}},
  \bibinfo{author}{\bibfnamefont{G.}~\bibnamefont{Andr\'e}},
  \bibinfo{author}{\bibfnamefont{M.}~\bibnamefont{Rams}},
  \bibinfo{author}{\bibfnamefont{K.}~\bibnamefont{Kr\'olas}},
  \bibinfo{author}{\bibfnamefont{C.}~\bibnamefont{Ritter}},
  \bibinfo{author}{\bibfnamefont{P.~C.~M.} \bibnamefont{Gubbens}},
  \bibnamefont{et~al.}, \bibinfo{journal}{Phys. Rev. Lett.}
  \textbf{\bibinfo{volume}{88}}, \bibinfo{pages}{077204}
  (\bibinfo{year}{2002}).

\bibitem[{\citenamefont{Ross et~al.}(2009)\citenamefont{Ross, Ruff, Adams,
  Gardner, Dabkowska, Qiu, Copley, and Gaulin}}]{Ross2009}
\bibinfo{author}{\bibfnamefont{K.~A.} \bibnamefont{Ross}},
  \bibinfo{author}{\bibfnamefont{J.~P.~C.} \bibnamefont{Ruff}},
  \bibinfo{author}{\bibfnamefont{C.~P.} \bibnamefont{Adams}},
  \bibinfo{author}{\bibfnamefont{J.~S.} \bibnamefont{Gardner}},
  \bibinfo{author}{\bibfnamefont{H.~A.} \bibnamefont{Dabkowska}},
  \bibinfo{author}{\bibfnamefont{Y.}~\bibnamefont{Qiu}},
  \bibinfo{author}{\bibfnamefont{J.~R.~D.} \bibnamefont{Copley}},
  \bibnamefont{and} \bibinfo{author}{\bibfnamefont{B.~D.}
  \bibnamefont{Gaulin}}, \bibinfo{journal}{Phys. Rev. Lett.}
  \textbf{\bibinfo{volume}{103}}, \bibinfo{pages}{227202}
  (\bibinfo{year}{2009}).

\bibitem[{\citenamefont{Ross et~al.}(2011{\natexlab{a}})\citenamefont{Ross,
  Yaraskavitch, Laver, Gardner, Quilliam, Meng, Kycia, Singh, Proffen,
  Dabkowska et~al.}}]{Ross2011}
\bibinfo{author}{\bibfnamefont{K.~A.} \bibnamefont{Ross}},
  \bibinfo{author}{\bibfnamefont{L.~R.} \bibnamefont{Yaraskavitch}},
  \bibinfo{author}{\bibfnamefont{M.}~\bibnamefont{Laver}},
  \bibinfo{author}{\bibfnamefont{J.~S.} \bibnamefont{Gardner}},
  \bibinfo{author}{\bibfnamefont{J.~A.} \bibnamefont{Quilliam}},
  \bibinfo{author}{\bibfnamefont{S.}~\bibnamefont{Meng}},
  \bibinfo{author}{\bibfnamefont{J.~B.} \bibnamefont{Kycia}},
  \bibinfo{author}{\bibfnamefont{D.~K.} \bibnamefont{Singh}},
  \bibinfo{author}{\bibfnamefont{T.}~\bibnamefont{Proffen}},
  \bibinfo{author}{\bibfnamefont{H.~A.} \bibnamefont{Dabkowska}},
  \bibnamefont{et~al.}, \bibinfo{journal}{Phys. Rev. B}
  \textbf{\bibinfo{volume}{84}}, \bibinfo{pages}{174442}
  (\bibinfo{year}{2011}{\natexlab{a}}).

\bibitem[{\citenamefont{Hermele et~al.}(2004)\citenamefont{Hermele, Fisher, and
  Balents}}]{Hermele2004}
\bibinfo{author}{\bibfnamefont{M.}~\bibnamefont{Hermele}},
  \bibinfo{author}{\bibfnamefont{M.~P.~A.} \bibnamefont{Fisher}},
  \bibnamefont{and} \bibinfo{author}{\bibfnamefont{L.}~\bibnamefont{Balents}},
  \bibinfo{journal}{Phys. Rev. B} \textbf{\bibinfo{volume}{69}},
  \bibinfo{pages}{064404} (\bibinfo{year}{2004}).

\bibitem[{\citenamefont{Gingras and McClarty}(2014)}]{GingrasRev}
\bibinfo{author}{\bibfnamefont{M.~J.~P.} \bibnamefont{Gingras}}
  \bibnamefont{and} \bibinfo{author}{\bibfnamefont{P.~A.}
  \bibnamefont{McClarty}}, \bibinfo{journal}{Reports on Progress in Physics}
  \textbf{\bibinfo{volume}{77}}, \bibinfo{pages}{056501}
  (\bibinfo{year}{2014}).

\bibitem[{\citenamefont{Bramwell et~al.}(2000)\citenamefont{Bramwell, Field,
  Harris, and Parkin}}]{Bramwell2000}
\bibinfo{author}{\bibfnamefont{S.~T.} \bibnamefont{Bramwell}},
  \bibinfo{author}{\bibfnamefont{M.~N.} \bibnamefont{Field}},
  \bibinfo{author}{\bibfnamefont{M.~J.} \bibnamefont{Harris}},
  \bibnamefont{and} \bibinfo{author}{\bibfnamefont{I.~P.}
  \bibnamefont{Parkin}}, \bibinfo{journal}{Journal of Physics: Condensed
  Matter} \textbf{\bibinfo{volume}{12}}, \bibinfo{pages}{483}
  (\bibinfo{year}{2000}).

\bibitem[{\citenamefont{Savary and Balents}(2012)}]{SavaryPRL}
\bibinfo{author}{\bibfnamefont{L.}~\bibnamefont{Savary}} \bibnamefont{and}
  \bibinfo{author}{\bibfnamefont{L.}~\bibnamefont{Balents}},
  \bibinfo{journal}{Phys. Rev. Lett.} \textbf{\bibinfo{volume}{108}},
  \bibinfo{pages}{037202} (\bibinfo{year}{2012}).

\bibitem[{\citenamefont{Savary and Balents}(2013)}]{SavaryPRB}
\bibinfo{author}{\bibfnamefont{L.}~\bibnamefont{Savary}} \bibnamefont{and}
  \bibinfo{author}{\bibfnamefont{L.}~\bibnamefont{Balents}},
  \bibinfo{journal}{Phys. Rev. B} \textbf{\bibinfo{volume}{87}},
  \bibinfo{pages}{205130} (\bibinfo{year}{2013}).

\bibitem[{\citenamefont{Ross et~al.}(2011{\natexlab{b}})\citenamefont{Ross,
  Savary, Gaulin, and Balents}}]{RossPRX}
\bibinfo{author}{\bibfnamefont{K.~A.} \bibnamefont{Ross}},
  \bibinfo{author}{\bibfnamefont{L.}~\bibnamefont{Savary}},
  \bibinfo{author}{\bibfnamefont{B.~D.} \bibnamefont{Gaulin}},
  \bibnamefont{and} \bibinfo{author}{\bibfnamefont{L.}~\bibnamefont{Balents}},
  \bibinfo{journal}{Phys. Rev. X} \textbf{\bibinfo{volume}{1}},
  \bibinfo{pages}{021002} (\bibinfo{year}{2011}{\natexlab{b}}).

\bibitem[{\citenamefont{Ross et~al.}(2012)\citenamefont{Ross, Proffen,
  Dabkowska, Quilliam, Yaraskavitch, Kycia, and Gaulin}}]{RossStuffing}
\bibinfo{author}{\bibfnamefont{K.~A.} \bibnamefont{Ross}},
  \bibinfo{author}{\bibfnamefont{T.}~\bibnamefont{Proffen}},
  \bibinfo{author}{\bibfnamefont{H.~A.} \bibnamefont{Dabkowska}},
  \bibinfo{author}{\bibfnamefont{J.~A.} \bibnamefont{Quilliam}},
  \bibinfo{author}{\bibfnamefont{L.~R.} \bibnamefont{Yaraskavitch}},
  \bibinfo{author}{\bibfnamefont{J.~B.} \bibnamefont{Kycia}}, \bibnamefont{and}
  \bibinfo{author}{\bibfnamefont{B.~D.} \bibnamefont{Gaulin}},
  \bibinfo{journal}{Phys. Rev. B} \textbf{\bibinfo{volume}{86}},
  \bibinfo{pages}{174424} (\bibinfo{year}{2012}).

\bibitem[{\citenamefont{Lau et~al.}(2006)\citenamefont{Lau, Muegge, McQueen,
  Duncan, and R.J.Cava}}]{Lau2006}
\bibinfo{author}{\bibfnamefont{G.~C.} \bibnamefont{Lau}},
  \bibinfo{author}{\bibfnamefont{B.}~\bibnamefont{Muegge}},
  \bibinfo{author}{\bibfnamefont{T.}~\bibnamefont{McQueen}},
  \bibinfo{author}{\bibfnamefont{E.}~\bibnamefont{Duncan}}, \bibnamefont{and}
  \bibinfo{author}{\bibnamefont{R.J.Cava}}, \bibinfo{journal}{J. Solid State
  Chem.} \textbf{\bibinfo{volume}{179}} (\bibinfo{year}{2006}).

\bibitem[{\citenamefont{Lau et~al.}(2008)\citenamefont{Lau, T.M.~McQueen, and
  Cava}}]{Lau2008}
\bibinfo{author}{\bibfnamefont{G.~C.} \bibnamefont{Lau}},
  \bibinfo{author}{\bibfnamefont{H.~Z.} \bibnamefont{T.M.~McQueen},
  \bibfnamefont{Q.Huang}}, \bibnamefont{and}
  \bibinfo{author}{\bibfnamefont{R.}~\bibnamefont{Cava}}, \bibinfo{journal}{J.
  Solid State Chem.} \textbf{\bibinfo{volume}{181}} (\bibinfo{year}{2008}).

\bibitem[{\citenamefont{Takatsu et~al.}(2012)\citenamefont{Takatsu, Kadowaki,
  Sato, Lynn, Tabata, Yamazaki, and Matsuhira}}]{Takatsu2012}
\bibinfo{author}{\bibfnamefont{H.}~\bibnamefont{Takatsu}},
  \bibinfo{author}{\bibfnamefont{H.}~\bibnamefont{Kadowaki}},
  \bibinfo{author}{\bibfnamefont{T.~J.} \bibnamefont{Sato}},
  \bibinfo{author}{\bibfnamefont{J.~W.} \bibnamefont{Lynn}},
  \bibinfo{author}{\bibfnamefont{Y.}~\bibnamefont{Tabata}},
  \bibinfo{author}{\bibfnamefont{T.}~\bibnamefont{Yamazaki}}, \bibnamefont{and}
  \bibinfo{author}{\bibfnamefont{K.}~\bibnamefont{Matsuhira}},
  \bibinfo{journal}{Journal of Physics: Condensed Matter}
  \textbf{\bibinfo{volume}{24}}, \bibinfo{pages}{052201}
  (\bibinfo{year}{2012}).

\bibitem[{\citenamefont{Yaouanc
  et~al.}(2011{\natexlab{a}})\citenamefont{Yaouanc, Dalmas~de R\'eotier,
  Chapuis, Marin, Vanishri, Aoki, F\aa{}k, Regnault, Buisson, Amato
  et~al.}}]{Yaouanc2011}
\bibinfo{author}{\bibfnamefont{A.}~\bibnamefont{Yaouanc}},
  \bibinfo{author}{\bibfnamefont{P.}~\bibnamefont{Dalmas~de R\'eotier}},
  \bibinfo{author}{\bibfnamefont{Y.}~\bibnamefont{Chapuis}},
  \bibinfo{author}{\bibfnamefont{C.}~\bibnamefont{Marin}},
  \bibinfo{author}{\bibfnamefont{S.}~\bibnamefont{Vanishri}},
  \bibinfo{author}{\bibfnamefont{D.}~\bibnamefont{Aoki}},
  \bibinfo{author}{\bibfnamefont{B.}~\bibnamefont{F\aa{}k}},
  \bibinfo{author}{\bibfnamefont{L.-P.} \bibnamefont{Regnault}},
  \bibinfo{author}{\bibfnamefont{C.}~\bibnamefont{Buisson}},
  \bibinfo{author}{\bibfnamefont{A.}~\bibnamefont{Amato}},
  \bibnamefont{et~al.}, \bibinfo{journal}{Phys. Rev. B}
  \textbf{\bibinfo{volume}{84}}, \bibinfo{pages}{184403}
  (\bibinfo{year}{2011}{\natexlab{a}}).

\bibitem[{\citenamefont{de~Reotier~et al.}(2006)}]{PalmasPhysicaB}
\bibinfo{author}{\bibfnamefont{P.~D.} \bibnamefont{de~Reotier~et al.}},
  \bibinfo{journal}{Physica B: Condensed Matter}
  \textbf{\bibinfo{volume}{374-375}}, \bibinfo{pages}{145}
  (\bibinfo{year}{2006}).

\bibitem[{\citenamefont{Blote et~al.}(1969)\citenamefont{Blote, Wielinga, and
  Huiskamp}}]{Blote}
\bibinfo{author}{\bibfnamefont{H.}~\bibnamefont{Blote}},
  \bibinfo{author}{\bibfnamefont{R.}~\bibnamefont{Wielinga}}, \bibnamefont{and}
  \bibinfo{author}{\bibfnamefont{W.}~\bibnamefont{Huiskamp}},
  \bibinfo{journal}{Physica} \textbf{\bibinfo{volume}{43}},
  \bibinfo{pages}{549} (\bibinfo{year}{1969}).

\bibitem[{\citenamefont{Yaouanc
  et~al.}(2011{\natexlab{b}})\citenamefont{Yaouanc, Dalmas~de R\'eotier, Marin,
  and Glazkov}}]{Yaouanc2011B}
\bibinfo{author}{\bibfnamefont{A.}~\bibnamefont{Yaouanc}},
  \bibinfo{author}{\bibfnamefont{P.}~\bibnamefont{Dalmas~de R\'eotier}},
  \bibinfo{author}{\bibfnamefont{C.}~\bibnamefont{Marin}}, \bibnamefont{and}
  \bibinfo{author}{\bibfnamefont{V.}~\bibnamefont{Glazkov}},
  \bibinfo{journal}{Phys. Rev. B} \textbf{\bibinfo{volume}{84}},
  \bibinfo{pages}{172408} (\bibinfo{year}{2011}{\natexlab{b}}).

\bibitem[{\citenamefont{Sala et~al.}(2014)\citenamefont{Sala, Gutmann,
  Prabhakaran, Pomaranski, Mitchelitis, Kycia, Porter, Castelnovo, and
  Goff}}]{Sala2014}
\bibinfo{author}{\bibfnamefont{G.}~\bibnamefont{Sala}},
  \bibinfo{author}{\bibfnamefont{M.~J.} \bibnamefont{Gutmann}},
  \bibinfo{author}{\bibfnamefont{D.}~\bibnamefont{Prabhakaran}},
  \bibinfo{author}{\bibfnamefont{D.}~\bibnamefont{Pomaranski}},
  \bibinfo{author}{\bibfnamefont{C.}~\bibnamefont{Mitchelitis}},
  \bibinfo{author}{\bibfnamefont{J.~B.} \bibnamefont{Kycia}},
  \bibinfo{author}{\bibfnamefont{D.~G.} \bibnamefont{Porter}},
  \bibinfo{author}{\bibfnamefont{C.}~\bibnamefont{Castelnovo}},
  \bibnamefont{and} \bibinfo{author}{\bibfnamefont{J.~P.} \bibnamefont{Goff}},
  \bibinfo{journal}{Nat Mater} \textbf{\bibinfo{volume}{13}},
  \bibinfo{pages}{488} (\bibinfo{year}{2014}).

\bibitem[{\citenamefont{Prather}(1961)}]{Prather}
\bibinfo{author}{\bibfnamefont{J.}~\bibnamefont{Prather}},
  \bibinfo{journal}{NBS monograph 19}  (\bibinfo{year}{1961}).

\bibitem[{\citenamefont{Stevens}(1952)}]{Stevens}
\bibinfo{author}{\bibfnamefont{K.~W.~H.} \bibnamefont{Stevens}},
  \bibinfo{journal}{Proceedings of the Physical Society. Section A}
  \textbf{\bibinfo{volume}{65}}, \bibinfo{pages}{209} (\bibinfo{year}{1952}).

\bibitem[{\citenamefont{Squires}(1978)}]{Squires}
\bibinfo{author}{\bibfnamefont{G.}~\bibnamefont{Squires}},
  \bibinfo{journal}{Cambridge University Press, Cambridge, UK}
  (\bibinfo{year}{1978}).

\bibitem[{\citenamefont{Granroth et~al.}(2010)\citenamefont{Granroth,
  Kolesnikov, Sherline, Clancy, Ross, Ruff, Gaulin, and Nagler}}]{Sequoia}
\bibinfo{author}{\bibfnamefont{G.~E.} \bibnamefont{Granroth}},
  \bibinfo{author}{\bibfnamefont{A.~I.} \bibnamefont{Kolesnikov}},
  \bibinfo{author}{\bibfnamefont{T.~E.} \bibnamefont{Sherline}},
  \bibinfo{author}{\bibfnamefont{J.~P.} \bibnamefont{Clancy}},
  \bibinfo{author}{\bibfnamefont{K.~A.} \bibnamefont{Ross}},
  \bibinfo{author}{\bibfnamefont{J.~P.~C.} \bibnamefont{Ruff}},
  \bibinfo{author}{\bibfnamefont{B.~D.} \bibnamefont{Gaulin}},
  \bibnamefont{and} \bibinfo{author}{\bibfnamefont{S.~E.}
  \bibnamefont{Nagler}}, \bibinfo{journal}{Journal of Physics: Conference
  Series} \textbf{\bibinfo{volume}{251}}, \bibinfo{pages}{012058}
  (\bibinfo{year}{2010}).

\bibitem[{\citenamefont{Bertin et~al.}(2012)\citenamefont{Bertin, Chapuis,
  de~R{\'e}otier, and Yaouanc}}]{Bertin2012}
\bibinfo{author}{\bibfnamefont{A.}~\bibnamefont{Bertin}},
  \bibinfo{author}{\bibfnamefont{Y.}~\bibnamefont{Chapuis}},
  \bibinfo{author}{\bibfnamefont{P.~D.} \bibnamefont{de~R{\'e}otier}},
  \bibnamefont{and} \bibinfo{author}{\bibfnamefont{A.}~\bibnamefont{Yaouanc}},
  \bibinfo{journal}{Journal of Physics: Condensed Matter}
  \textbf{\bibinfo{volume}{24}}, \bibinfo{pages}{256003}
  (\bibinfo{year}{2012}).

\bibitem[{\citenamefont{Cao et~al.}(2009)\citenamefont{Cao, Gukasov, Mirebeau,
  and Bonville}}]{Cao2009}
\bibinfo{author}{\bibfnamefont{H.~B.} \bibnamefont{Cao}},
  \bibinfo{author}{\bibfnamefont{A.}~\bibnamefont{Gukasov}},
  \bibinfo{author}{\bibfnamefont{I.}~\bibnamefont{Mirebeau}}, \bibnamefont{and}
  \bibinfo{author}{\bibfnamefont{P.}~\bibnamefont{Bonville}},
  \bibinfo{journal}{Journal of Physics: Condensed Matter}
  \textbf{\bibinfo{volume}{21}}, \bibinfo{pages}{492202}
  (\bibinfo{year}{2009}).

\bibitem[{\citenamefont{Hodges et~al.}(2001)\citenamefont{Hodges, Bonville,
  Forget, Rams, Kr{\'o}las, and Dhalenne}}]{Hodges2001}
\bibinfo{author}{\bibfnamefont{J.~A.} \bibnamefont{Hodges}},
  \bibinfo{author}{\bibfnamefont{P.}~\bibnamefont{Bonville}},
  \bibinfo{author}{\bibfnamefont{A.}~\bibnamefont{Forget}},
  \bibinfo{author}{\bibfnamefont{M.}~\bibnamefont{Rams}},
  \bibinfo{author}{\bibfnamefont{K.}~\bibnamefont{Kr{\'o}las}},
  \bibnamefont{and} \bibinfo{author}{\bibfnamefont{G.}~\bibnamefont{Dhalenne}},
  \bibinfo{journal}{Journal of Physics: Condensed Matter}
  \textbf{\bibinfo{volume}{13}}, \bibinfo{pages}{9301} (\bibinfo{year}{2001}).

\bibitem[{\citenamefont{Malkin et~al.}(2004)\citenamefont{Malkin, Zakirov,
  Popova, Klimin, Chukalina, Antic-Fidancev, Goldner, Aschehoug, and
  Dhalenne}}]{Malkin2004}
\bibinfo{author}{\bibfnamefont{B.~Z.} \bibnamefont{Malkin}},
  \bibinfo{author}{\bibfnamefont{A.~R.} \bibnamefont{Zakirov}},
  \bibinfo{author}{\bibfnamefont{M.~N.} \bibnamefont{Popova}},
  \bibinfo{author}{\bibfnamefont{S.~A.} \bibnamefont{Klimin}},
  \bibinfo{author}{\bibfnamefont{E.~P.} \bibnamefont{Chukalina}},
  \bibinfo{author}{\bibfnamefont{E.}~\bibnamefont{Antic-Fidancev}},
  \bibinfo{author}{\bibfnamefont{P.}~\bibnamefont{Goldner}},
  \bibinfo{author}{\bibfnamefont{P.}~\bibnamefont{Aschehoug}},
  \bibnamefont{and} \bibinfo{author}{\bibfnamefont{G.}~\bibnamefont{Dhalenne}},
  \bibinfo{journal}{Phys. Rev. B} \textbf{\bibinfo{volume}{70}},
  \bibinfo{pages}{075112} (\bibinfo{year}{2004}).

\bibitem[{\citenamefont{Hutchings}(1964)}]{Hutchings}
\bibinfo{author}{\bibfnamefont{M.}~\bibnamefont{Hutchings}}, in
  \emph{\bibinfo{booktitle}{Solid State Physics}}, edited by
  \bibinfo{editor}{\bibfnamefont{F.}~\bibnamefont{Seitz}} \bibnamefont{and}
  \bibinfo{editor}{\bibfnamefont{D.}~\bibnamefont{Turnbull}}
  (\bibinfo{publisher}{Academic Press}, \bibinfo{year}{1964}),
  vol.~\bibinfo{volume}{16}, pp. \bibinfo{pages}{227 -- 273}.

\bibitem[{\citenamefont{Walter}(1984)}]{Walter1984}
\bibinfo{author}{\bibfnamefont{U.}~\bibnamefont{Walter}},
  \bibinfo{journal}{Journal of Physics and Chemistry of Solids}
  \textbf{\bibinfo{volume}{45}}, \bibinfo{pages}{401 } (\bibinfo{year}{1984}),
  ISSN \bibinfo{issn}{0022-3697}.

\bibitem[{\citenamefont{Freeman and Watson}(1962)}]{Freeman1962}
\bibinfo{author}{\bibfnamefont{A.~J.} \bibnamefont{Freeman}} \bibnamefont{and}
  \bibinfo{author}{\bibfnamefont{R.~E.} \bibnamefont{Watson}},
  \bibinfo{journal}{Phys. Rev.} \textbf{\bibinfo{volume}{127}},
  \bibinfo{pages}{2058} (\bibinfo{year}{1962}).

\bibitem[{\citenamefont{Siddharthan et~al.}(1999)\citenamefont{Siddharthan,
  Shastry, Ramirez, Hayashi, Cava, and Rosenkranz}}]{Siddhartan1999}
\bibinfo{author}{\bibfnamefont{R.}~\bibnamefont{Siddharthan}},
  \bibinfo{author}{\bibfnamefont{B.~S.} \bibnamefont{Shastry}},
  \bibinfo{author}{\bibfnamefont{A.~P.} \bibnamefont{Ramirez}},
  \bibinfo{author}{\bibfnamefont{A.}~\bibnamefont{Hayashi}},
  \bibinfo{author}{\bibfnamefont{R.~J.} \bibnamefont{Cava}}, \bibnamefont{and}
  \bibinfo{author}{\bibfnamefont{S.}~\bibnamefont{Rosenkranz}},
  \bibinfo{journal}{Phys. Rev. Lett.} \textbf{\bibinfo{volume}{83}},
  \bibinfo{pages}{1854} (\bibinfo{year}{1999}).

\bibitem[{\citenamefont{Arnold et~al.}(2014)\citenamefont{Arnold, Bilheux,
  Borreguero, Buts, Campbell, Chapon, Doucet, Draper, Leal, Gigg
  et~al.}}]{Mantid}
\bibinfo{author}{\bibfnamefont{O.}~\bibnamefont{Arnold}},
  \bibinfo{author}{\bibfnamefont{J.}~\bibnamefont{Bilheux}},
  \bibinfo{author}{\bibfnamefont{J.}~\bibnamefont{Borreguero}},
  \bibinfo{author}{\bibfnamefont{A.}~\bibnamefont{Buts}},
  \bibinfo{author}{\bibfnamefont{S.}~\bibnamefont{Campbell}},
  \bibinfo{author}{\bibfnamefont{L.}~\bibnamefont{Chapon}},
  \bibinfo{author}{\bibfnamefont{M.}~\bibnamefont{Doucet}},
  \bibinfo{author}{\bibfnamefont{N.}~\bibnamefont{Draper}},
  \bibinfo{author}{\bibfnamefont{R.}~\bibnamefont{Leal}},
  \bibinfo{author}{\bibfnamefont{M.}~\bibnamefont{Gigg}}, \bibnamefont{et~al.},
  \bibinfo{journal}{Nuclear Instruments and Methods in Physics Research Section
  A: Accelerators, Spectrometers, Detectors and Associated Equipment}
  \textbf{\bibinfo{volume}{764}}, \bibinfo{pages}{156} (\bibinfo{year}{2014}).

\bibitem[{\citenamefont{Azuah et~al.}(2009)\citenamefont{Azuah, Kneller, Qiu,
  Brown, Copley, Dimeo, and Tregenna-Piggott}}]{Dave}
\bibinfo{author}{\bibfnamefont{R.}~\bibnamefont{Azuah}},
  \bibinfo{author}{\bibfnamefont{L.}~\bibnamefont{Kneller}},
  \bibinfo{author}{\bibfnamefont{Y.}~\bibnamefont{Qiu}},
  \bibinfo{author}{\bibfnamefont{C.}~\bibnamefont{Brown}},
  \bibinfo{author}{\bibfnamefont{J.}~\bibnamefont{Copley}},
  \bibinfo{author}{\bibfnamefont{R.}~\bibnamefont{Dimeo}}, \bibnamefont{and}
  \bibinfo{author}{\bibfnamefont{P.}~\bibnamefont{Tregenna-Piggott}},
  \bibinfo{journal}{J. Res. Natl. Inst. Stan. Technol.}
  \textbf{\bibinfo{volume}{114}} (\bibinfo{year}{2009}).

\end{thebibliography}

\end{document}